\PassOptionsToPackage{merge}{natbib}
\documentclass[epj]{webofc}
\usepackage[varg]{txfonts}   % Web of Conferences font
\usepackage{subcaption}
\usepackage{bm}
\wocname{EPJ Web of Conferences}
\woctitle{Winter Workshop on Non-Perturbative Quantum Field Theory}
\newcommand{\cN}{{\cal N}}

\newcommand{\cD}{{\cal D}}

\newcommand{\be}{\begin{equation}}
\newcommand{\beq}{\begin{equation}}
\newcommand{\ee}{\end{equation}}
\newcommand{\eeq}{\end{equation}}
\newcommand{\bea}{\begin{eqnarray}}
\newcommand{\eea}{\end{eqnarray}}

\newcommand{\rkx}{\right)}
\newcommand{\lk}{\left(}

\newcommand{\sli}{\sum\limits}

\newcommand{\il}{\int\limits}

\def\R{{\mathbb{R}}}

\newcommand{\va}{\vec{a}}
\newcommand{\vB}{\vec{B}}
\newcommand{\ve}{\vec{e}}
\newcommand{\vA}{\vec{A}}
\newcommand{\vx}{\vec{x}}
\newcommand{\vy}{\vec{y}}
\newcommand{\vz}{\vec{z}}
\newcommand{\vD}{\vec{D}}
\newcommand{\vk}{\vec{k}}
\newcommand{\vp}{\vec{p}}
\newcommand{\vq}{\vec{q}}

\newcommand{\vE}{\vec{E}}

\newcommand{\ii}{\mathrm{i}}
\newcommand{\dd}{\mathrm{d}}
\newcommand{\tr}{\mathrm{tr}}

\renewcommand{\vec}[1]{\bm{#1}}

\begin{document}
\selectlanguage{english}
\title{Hamiltonian approach to QCD in Coulomb gauge -- a survey of recent results\footnote{Talk given by H.~Reinhardt at "5th Winter Workshop on Non-Perturbative Quantum Field Theory, 22-24 March 2017, Sophia-Antipolis, France}}
%
% subtitle (optional, strongly discouraged)

\author{H.~Reinhardt\inst{1}\fnsep\thanks{\email{hugo.reinhardt@uni-tuebingen.de}} \and
	G.~Burgio\inst{1} \and D.~Campagnari\inst{1} \and E.~Ebadati\inst{1} \and J.~Heffner\inst{1} \and M.~Quandt\inst{1} \and P.~Vastag\inst{1} \and H.~Vogt\inst{1}
}

\institute{Universit\"at T\"ubingen \\
Institut f\"ur Theoretische Physik \\
Auf der Morgenstelle 14 \\
D-72076 T\"ubingen \\
Germany}

\abstract{I report on recent results obtained within the Hamiltonian approach to QCD in Coulomb gauge. Furthermore this approach is compared to recent lattice data, which were obtained by an alternative gauge fixing method and which show an improved agreement with the continuum results. By relating the Gribov confinement scenario to the center vortex picture of confinement it is shown that the Coulomb string tension is tied to the spatial string tension. For the quark sector a vacuum wave functional is used which explicitly contains the coupling of the quarks to the transverse gluons and which results in variational equations which are free of ultraviolet divergences. The variational approach is extended to finite temperatures by compactifying a spatial dimension. The effective potential of the Polyakov loop is evaluated from the zero-temperature variational solution. For pure Yang--Mills theory, the deconfinement phase transition is found to be second order for SU(2) and first order for SU(3), in agreement with the lattice results. The corresponding  critical temperatures are found to be $275 \, \mathrm{MeV}$ and $280 \, \mathrm{MeV}$, respectively. When quarks are included, the deconfinement transition turns into a cross-over. From the dual and chiral quark condensate one finds pseudo-critical temperatures of $198 \, \mathrm{MeV}$ and $170 \, \mathrm{MeV}$, respectively, for the deconfinement and chiral transition.
}
\maketitle
\section{Introduction}
\label{intro}

One of the most challenging problems in particle physics is the understanding of the phase diagram of strongly interacting matter. By means of ultra-relativistic heavy ion collisions the properties of hadronic matter at high temperature and/or density can be explored. From the theoretical point of view we have access to the finite-temperature behavior of QCD by means of lattice Monte-Carlo calculations. Due to the sign problem, this method fails, however, to describe baryonic matter at high density or, more technically, QCD at large chemical baryon potential \cite{Gattringer2016}. Therefore, alternative, non-perturbative approaches to QCD which do not rely on the lattice formulation and hence do not suffer from the notorious sign problem are desirable. In recent years, much effort has been devoted to develop non-perturbative continuum approaches. These are based either on Dyson--Schwinger equations \cite{Fischer2006, Alkofer2001, Binosi2009, Watson2006, Watson2007, Watson2008} or functional renormalization group flow equations \cite{Pawlowski2007, Gies2012}, or they exploit the variational principle in either the Hamiltonian \cite{Feuchter2004, Feuchter2005} or covariant \cite{Quandt2013, Quandt2015} formulation of gauge theory. There are also semiphenomenological approaches assuming a massive gluon propagator \cite{Reinosa:2014zta,*RSTW2016} or the Gribov--Zwanziger action \cite{Gribov1978,*Zwanziger:1988jt,*Zwanziger:1989mf,*Zwanziger:1992qr}, see Ref.~\cite{Canfora2015}. 

In this talk, I will review some recent results obtained within the Hamiltonian approach to QCD in Coulomb gauge both at zero and finite temperatures. After a short introduction to the basic features of this approach I will summarize the essential zero-temperature results for pure Yang--Mills theory and compare
them to recent lattice data which were obtained 
by an alternative gauge fixing method, which is expected to yield results closer to the continuum theory. After that, I will show by means of lattice calculations that the so-called Coulomb string tension is linked not to the temporal but to the spatial string tension. In this context, I will demonstrate that the Gribov--Zwanziger confinement scenario is related to the center vortex picture of confinement. I will then report on new variational calculations carried out for the quark sector of QCD. After that I will extend the Hamiltonian approach to QCD in Coulomb gauge to finite temperatures by compactifying a spatial dimension \cite{Reinhardt:2016xci}. Numerical results will be given for the Polyakov loop and the chiral and dual quark condensates. Finally, I will give some outlook on future research within the Hamiltonian approach.

\section{Variational Hamiltonian approach to Yang--Mills theory}\label{sectII}

For pedagogical reason let me first summarize the basic features of the Hamiltonian approach in Coulomb gauge for pure Yang--Mills theory.
The Hamiltonian approach to Yang--Mills theory starts from Weyl gauge $A_0(\vx) = 0$ and considers the spatial components of the gauge field $A_i^a(\vx)$ as coordinates. The momenta are introduced in the standard fashion $\Pi_i^a(\vx) = \delta S_{\mathrm{YM}}[A] / \delta \dot{A}_i^a(\vx) = -E_i^a(\vx)$ and turn out to be the color electric field $\vE^a(\vx)$. The classical Yang--Mills Hamiltonian is then obtained as
\beq
H = \frac{1}{2} \int d^3 x \left( \vec{E}^2(\vx) + \vB^2(\vx) \right) \, , \label{94-1}
\eeq
where
\beq
B^a_k(\vx) = \varepsilon_{klm} \left(\partial_l A^a_m(\vx) - \frac{g}{2} f^{abc} A^b_l(\vx) A^c_m(\vx)\right)
\eeq
is the non-Abelian color magnetic field with $g$ being the coupling constant. The theory is quantized by replacing the classical momentum $\Pi_i^a$ by the operator $\Pi_i^a(\vx) = -\ii \delta / \delta A_i^a(\vx)$. The central issue is then to solve the Schr\"odinger equation $H \phi[A] = E \phi[A]$ for the vacuum wave functional $\phi[A]$. Due to the use of Weyl gauge, Gau{\ss}'s law $\hat{\vD} \cdot \vec{\Pi} \phi[A] = 0$ (with $\hat{\vD} = \vec{\partial} + g \vA$ being the covariant derivative in the adjoint representation) has to be put as a constraint on the wave functional, which ensures the gauge invariance of the latter. Instead of working with explicitly gauge invariant states, it is more convenient to fix
the gauge and explicitly resolve Gau{\ss}'s law in the chosen gauge. This has the advantage that 
any (normalizable) wave functional $\phi[A]$ is physically admissable for a variational approach, 
while the price to pay is a significant complication of the gauge-fixed Hamiltonian. 
A particular convenient choice of gauge for this method turns out to be 
Coulomb gauge $\vec{\partial} \cdot \vec{A} = 0$. 

After canonical quantization in Weyl gauge $A_0 = 0$ and resolution of Gau{\ss}'s law in Coulomb gauge $\vec{\partial} \cdot \vA = 0$ one finds the following gauge fixed Hamiltonian \cite{Christ1980}
\beq
H = H_{\mathrm{T}} + H_{\mathrm{C}} \label{232-1}
\eeq
with
\beq
H_{\mathrm{T}} = \frac{1}{2} \int \dd^3 x \lk J^{-1}[A] \vec{\Pi}^a(\vx) \cdot J[A] \vec{\Pi}^a(\vx) + \vB^a(\vx) \cdot \vB^a(\vx) \right) \label{G1}\,,
\eeq
where
\beq
J[A] = \mathrm{Det}(-\hat{\vD} \cdot \vec{\partial}) \label{203-1}
\eeq
is the Fadeev--Popov determinant and
\beq
H_{\mathrm{C}} = \frac{g^2}{2} \int \dd^3 x \int \dd^3 y \, J[A]^{-1} \rho^a(\vx) J[A] \left[(-\hat{\vD} \cdot \vec{\partial})^{-1} (-\vec{\partial}^2) (-\hat{\vD} \cdot \vec{\partial})^{-1}\right]^{a b}\!\!(\vx, \vy) \, \rho^b(\vy) \label{2}
\eeq
is the so-called Coulomb term with the color charge density
\beq
\rho^a(\vx) = f^{a b c} \vA^b(\vx) \cdot \vec{\Pi}^c(\vx) + \rho_m^a(\vx)\,. \label{3}
\eeq
This expression contains besides the charge density of the matter fields $\rho^a_m$ also a purely gluonic part. Due to the implementation of Coulomb gauge, the scalar product in the Hilbert space of wave functionals $\phi[A] = \langle A \vert \phi \rangle$ is defined by
\beq
\langle \phi | \ldots | \psi \rangle = \int \cD A \, J[A] \,\phi^*[A] \ldots \psi[A] \, . \label{414-G4}
\eeq
Here, the functional integration is over transversal spatial gauge fields and the Fadeev--Popov determinant $J[A]$ appears due to Coulomb gauge fixing with the standard Fadeev--Popov method. The Faddeev--Popov determinant (\ref{203-1}) in the integration measure represents the Jacobian of the change of variables from ``Cartesian'' to ``curvilinear'' variables 
in Coulomb gauge. With the gauge fixed Hamiltonian (\ref{232-1}) one has to solve the stationary Schr\"odinger equation $H \phi[A] = E \phi[A]$ for the vacuum wave functional $\phi[A]$. Once $\phi[A]$ is known, all observables and correlation functions can, in principle, be calculated. This has been attempted by means of the variational principle 
using Gaussian type ans\"atze for the vacuum wave functional \cite{Schuette1984,SS2001}. However, the first attempts did not properly include the Faddeev--Popov determinant, which turns out to be  crucial in order to describe the confinement properties of the theory. Below, I will discuss the variational approach developed in Refs.~\cite{Feuchter2004,Feuchter2005}, which differs from previous attempts by the ansatz for the vacuum wave functional, the treatment of the Faddeev--Popov determinant and, equally important, by the renormalization; see Ref.~\cite{Greensite2011a} for further details.

\subsection{Variational solution of the Schr\"odinger equation}

The ansatz for the vacuum wave functional is inspired by the quantum mechanics of a  particle in a spherically symmetric potential for which the ground state wave function is given by $\phi(r) = u(r) / r$, where the radial wave functional $u(r)$ satisfies a standard one-dimensional Schr\"odinger equation and $r$ represents (the square root of the radial part of) the Jacobian of the transformation from the Cartesian to spherical coordinates. Our ansatz for the vacuum wave functional is given by
\beq
\phi_{\mathrm{YM}}[A] = \frac{1}{\sqrt{J[A]}} \exp\left[-\frac{1}{2} \int \dd^3 x \int \dd^3 y \, A_k^a(\vx) \omega(\vx, \vy) A_k^a(\vy)\right] \equiv \frac{1}{\sqrt{J [A]}} \, \tilde{\phi}_{\mathrm{YM}}[A] \, . \label{G6}
\eeq
The inclusion of the pre-exponential factor has the advantage that it eliminates the Faddeev--Popov determinant from the integration measure in the scalar product (\ref{414-G4}). Furthermore, for the wave function (\ref{G6}) the gluon propagator is given up to a factor of $\frac{1}{2}$ by the inverse of the variational kernel $\omega(\vx, \vy)$. It turns out 
that in the Yang--Mills sector the Coulomb term $H_{\mathrm{C}}$ (\ref{2}) can be ignored.

\begin{figure}
\centering
\includegraphics[width=5cm,clip]{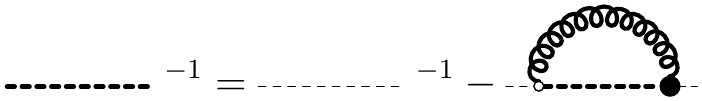}
\caption{Dyson--Schwinger equation for the ghost propagator.}
\label{fig1} %
\end{figure}

Calculating the expectation value of the remaining parts of the Yang--Mills Hamiltonian (\ref{G1}) with the wave functional (\ref{G6}) up to two loops, the minimization of the energy density with respect to $\omega(\vx, \vy)$ yields the following gap equation in momentum space\footnote{Due to translational and rotational invariance, kernels such as $\omega(\vx, \vy)$ can be Fourier transformed as
\[
\omega(\vx, \vy) = \int \frac{\dd^3 k}{(2 \pi)^3}\,\mathrm{e}^{\ii \vk \cdot (\vx - \vy)}\,\omega(k) \, ,
\] 
where the new kernel in momentum space depends on $k = |\vk|$ only. For simplicity, we will use the same symbol for the kernel in position and momentum space and go back and forth between both representations with impunity.}
\beq
\omega^2(k) = \vk^2 + \chi^2(k) + c \, , \label{G8}
\eeq
where $c$ is a finite renormalization constant resulting from the tadpole and
\beq
\chi_{k l}^{a b}(\vx, \vy) = -\frac{1}{2} \Bigl\langle \phi \Big\vert \frac{\delta^2 \ln J[A]}{\delta A_k^a(\vx) \delta A_l^b(\vy)} \Big\vert \phi \Bigr\rangle = \delta^{ab} t_{kl} (\vx - \vy) \chi (\vx - \vy) \label{G9}
\eeq
represents the ghost loop. This can be expressed in terms of the ghost propagator
\beq
G(\vx, \vy) = \langle \phi \vert {\bigl(-\hat{\vD} \cdot \vec{\partial}\bigr)^{-1}}(\vx, \vy) \vert \phi \rangle \, , \label{240}
\eeq
which is evaluated with the vacuum wave functional (\ref{G6}), in an approximate way, 
resulting in a Dyson--Schwinger equation for the form factor
\beq
d(\vk) = g \vk^2 G(\vk) \label{246}
\eeq
of the ghost propagator which is diagrammatically illustrated in fig.~\ref{fig1}. This equation has to be solved together with the gap equation (\ref{G8}).

Dyson--Schwinger equations are functional differential equations and their solutions are uniquely determined only after providing appropriate boundary conditions. In the present case, the so-called horizon condition
\beq
d^{-1}(0) = 0 \label{272GX}
\eeq
is assumed, which is the key point in Gribov's confinement scenario. Its physical implications will be discussed later.

The equations given in eq.~(\ref{G8}) and fig.~\ref{fig1} can be solved analytically in the infrared using power law ans\"atze
\beq
\omega(p) = A p^{-\alpha} \, , \qquad d(p) = B p^{-\beta} \, . \label{308-x1}
\eeq
Assuming a bare ghost-gluon vertex and the horizon condition (\ref{272GX}), one finds for the IR exponents of gluon and ghost form factor (\ref{308-x1}) the sum rule
\beq
\alpha = 2 \beta - (d - 2) \, , \label{314-x2}
\eeq
where $d$ is the number of spatial dimensions (i.e.~$d = 3$ is our real world). The coupled gluon gap equation (\ref{G8}) and ghost DSE (Fig.~\ref{fig1}) allow for a single solution in $d = 2$,
\beq
\beta = 0.4 \, , \label{320-x3}
\eeq
and for two solutions in $d = 3$,
\beq
\beta = 1 \, \qquad \mbox{and} \qquad \beta = 0.796 \, . \label{325-x4}
\eeq

\begin{figure}
\centering
\includegraphics[width=7cm,clip]{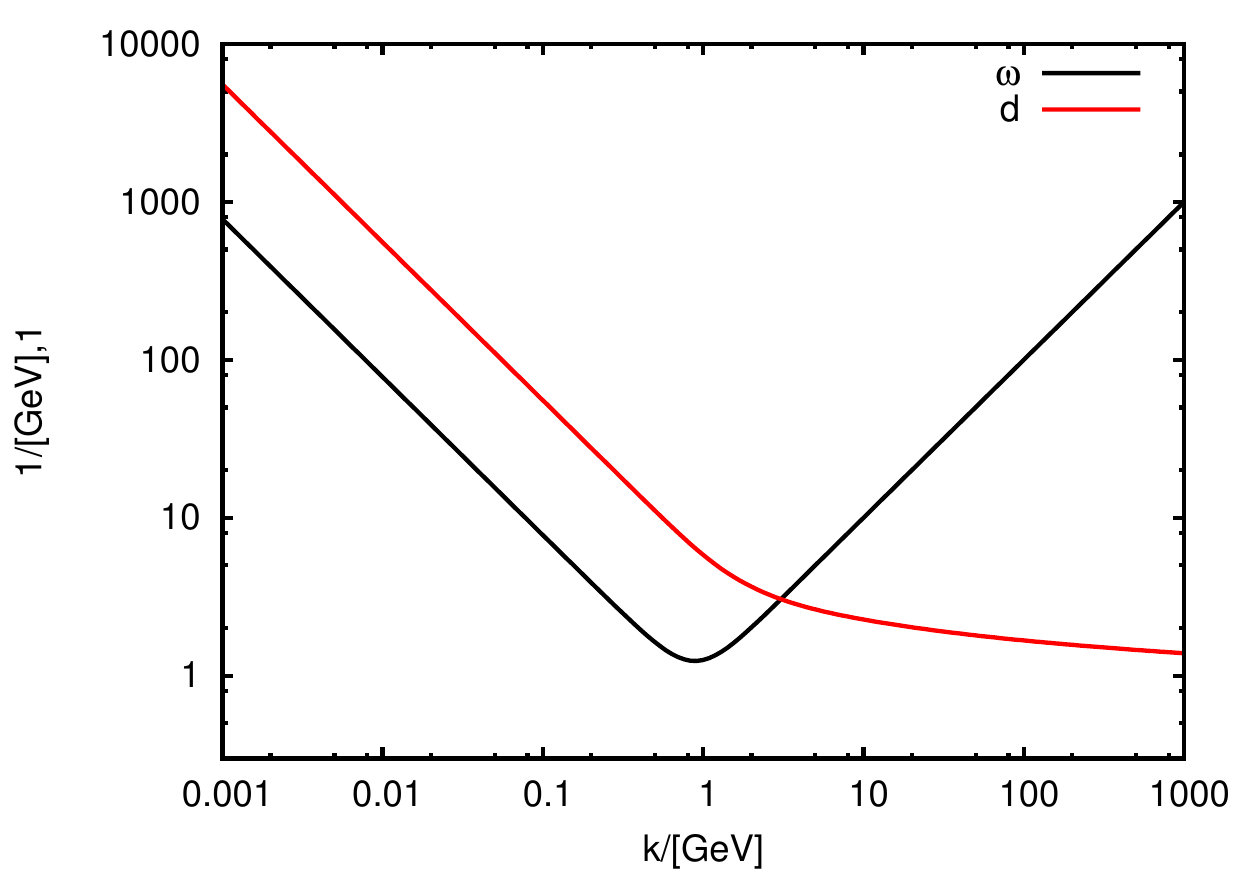}
\caption{Numerical solution of the coupled gap equation for $\omega$ (\ref{G8}) and Dyson--Schwinger equation for the ghost form factor $d$ (\ref{246}) for the renormalization constant $c = 0$ for $d = 3$ spatial dimensions \cite{ERS2007}.}
\label{fig2} %
\end{figure}

\clearpage
The numerical solutions of the gluon gap and ghost DSE are shown in fig.~\ref{fig2}. The numerical solutions reproduce the result (\ref{325-x4}) of the IR analysis. At large momenta the gluon energy $\omega(p)$ approaches the photon energy $|\vp|$ in agreement with asymptotic freedom, while $\omega(p)$ diverges like $\sim 1 / |\vp|$ in the IR, which implies the absence of free gluons in the IR and signals confinement.

\begin{figure}
\centering
\begin{subfigure}{0.45\textwidth}
\includegraphics[width=\textwidth,clip]{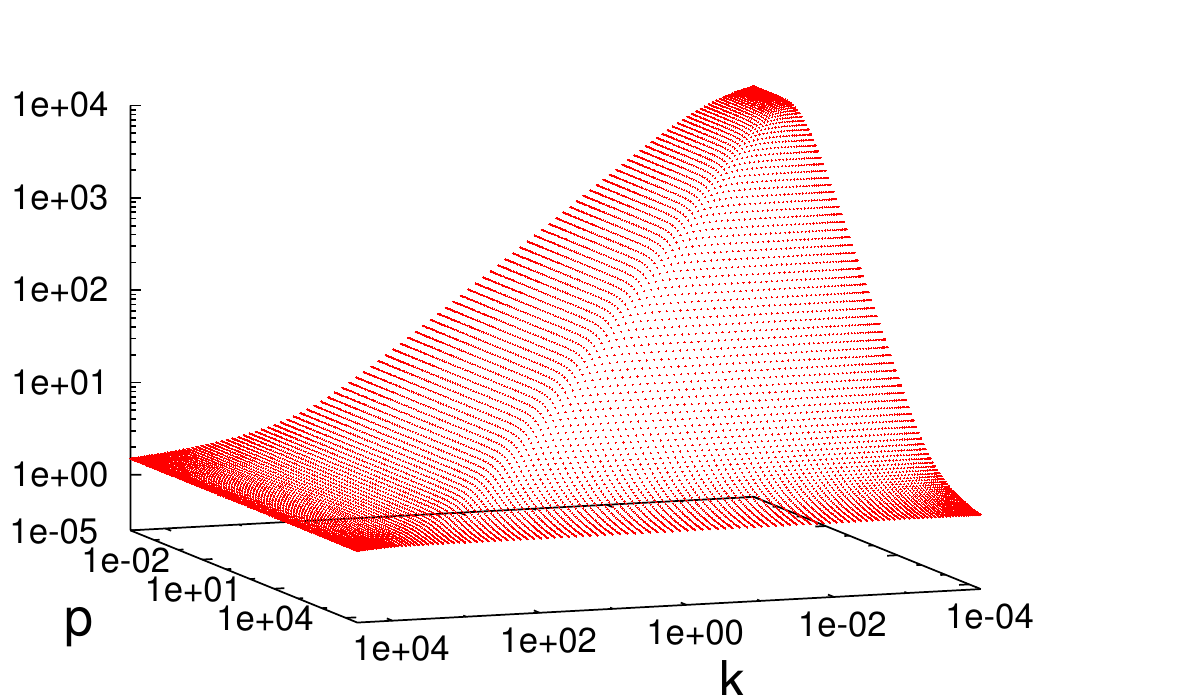}
\caption{}
\end{subfigure}
\quad
\begin{subfigure}{0.45\textwidth}
\includegraphics[width=\textwidth,clip]{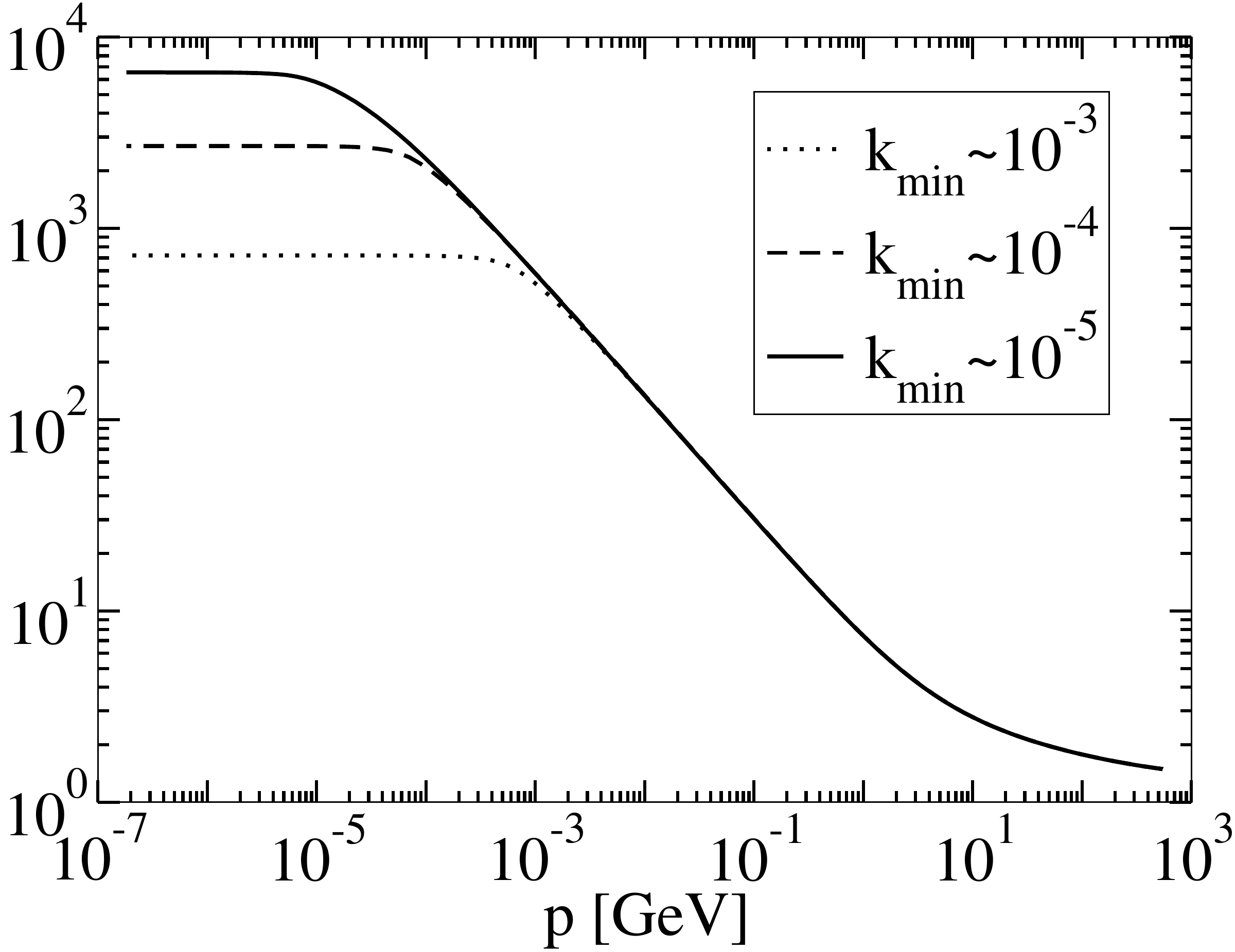}
\caption{}
\end{subfigure}
\caption{(a) The ghost form factor obtained in Ref.~\cite{Leder2011} from the solution of the renormalization group flow equations. Here, $p$ represents the momentum variable of the ghost form factor while $k$ is the infrared momentum cutoff of the flow equations. (b) Cuts through subfigure (a) at various values of the momentum scale $k$ of the flow equations.}
\label{fig3}%
\end{figure}

Alternatively to the variational approach, one can indirectly determine the vacuum wave functional by solving the functional renormalization group flow equations for the various propagators and vertex  functions of the Hamiltonian approach. Restricting the flow equations to those for the ghost and gluon propagators, one finds for the ghost form factor the result shown in fig.~\ref{fig3}. Starting with a constant ghost form factor in the ultraviolet, the ghost form factor develops an infrared singularity as the momentum cutoff of the flow equation tends to zero. This is nicely seen in fig.~\ref{fig3} (b), which shows a cut through fig.~\ref{fig3} (a) at fixed renormalization group scale $k$.

Let us also mention that it is not necessary to assume the horizon condition (\ref{272GX}) in the case of $D = 2 + 1$ dimensions, where it is a direct consequence of the coupled equations for the ghost and gluon propagators obtained from the variational principle. Finally, the horizon condition (\ref{272GX}) is also seen in the lattice data for the ghost form factor, see section \ref{sect3}.

\subsection{Physical implications of the ghost form factor}

As can be seen from its definition (\ref{246}), the ghost form factor expresses the deviation of Yang--Mills theory from QED, where the Faddeev--Popov operator in Coulomb gauge is given by the Laplacian, i.e. the ghost propagator is $G (p) = 1/p^2$.

Coulomb gauge is called a physical gauge since in QED the remaining transversal components are the gauge invariant degrees of freedom. This is not the case for Yang--Mills theory. However, Coulomb gauge can be viewed as a physical gauge also in the case of Yang--Mills theory in the sense that the inverse ghost form factor in Coulomb gauge represents the dielectric function of the Yang--Mills vacuum \cite{Reinhardt2008}
\beq
\epsilon(k) = d^{-1}(k) \, . \label{G12}
\eeq
The horizon condition (\ref{272GX}) guarantees that this function vanishes in the infrared, $\epsilon(k = 0) = 0$. This implies that the Yang--Mills vacuum is a perfect color dielectricum, i.e.~a dual superconductor. In this way the Hamiltonian approach in Coulomb gauge relates Gribov's confinement scenario to the dual Mei{\ss}ner effect, a confinement mechanism realized through the condensation of magnetic monopoles and proposed by Mandelstam and 't Hooft \cite{Mandelstam:1974pi,tHooft:1982ylj}. The dielectric function obtained here as inverse ghost form factor is also in accord with the phenomenological bag model picture of hadrons. Inside the hadron, i.e.~at small distance, the dielectric function is that of a normal vacuum while outside the physical hadrons the vanishing of the dielectric constant implies the absence of free color charges by Gau{\ss}'s law.

\section{Comparison with lattice calculation} \label{sect3}

\begin{figure}
\centering
\begin{subfigure}{0.35\textwidth}
\includegraphics[width=\textwidth,clip]{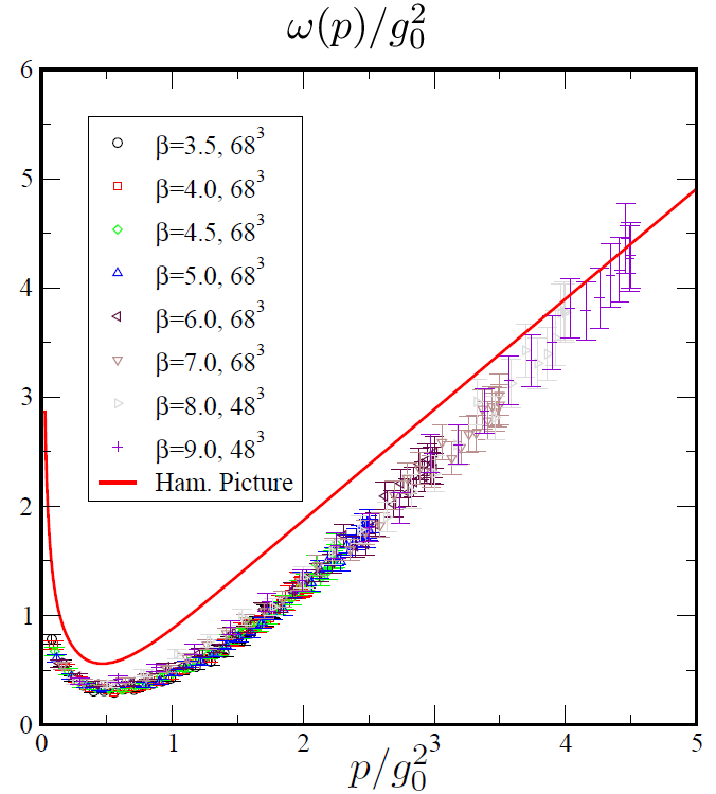}
\caption{}
\end{subfigure}
\hspace{0.17\textwidth}
\begin{subfigure}{0.33\textwidth}
\includegraphics[width=\textwidth,clip]{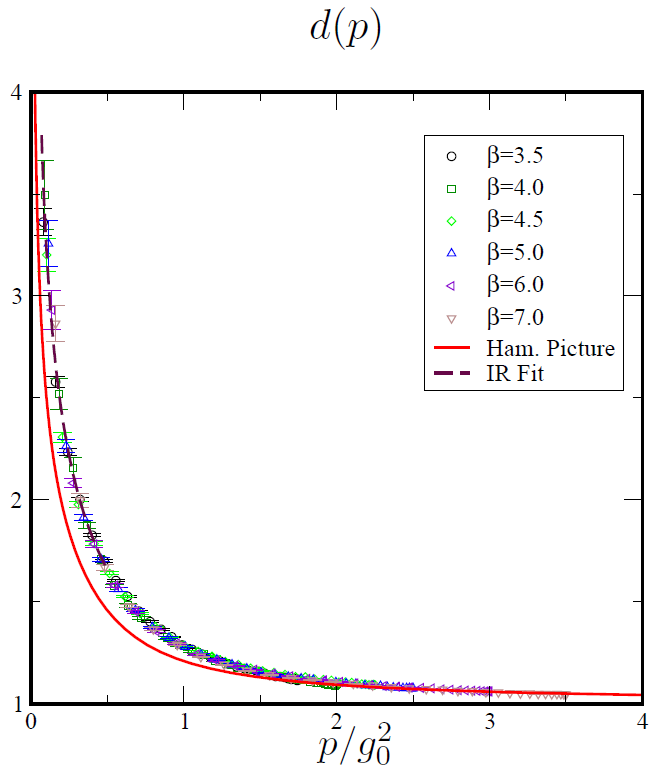}
\caption{}
\end{subfigure}
\caption{Comparison of the variational approach to $(2 + 1)$-dimensional Yang--Mills theory in Coulomb gauge \cite{Feuchter:2007mq} with the lattice data \cite{Moyaerts_thesis}: (a) gluon energy and (b) ghost form factor.}
\label{figX}%
\end{figure}%

\begin{figure}
\centering
\includegraphics[width=0.45\textwidth,clip]{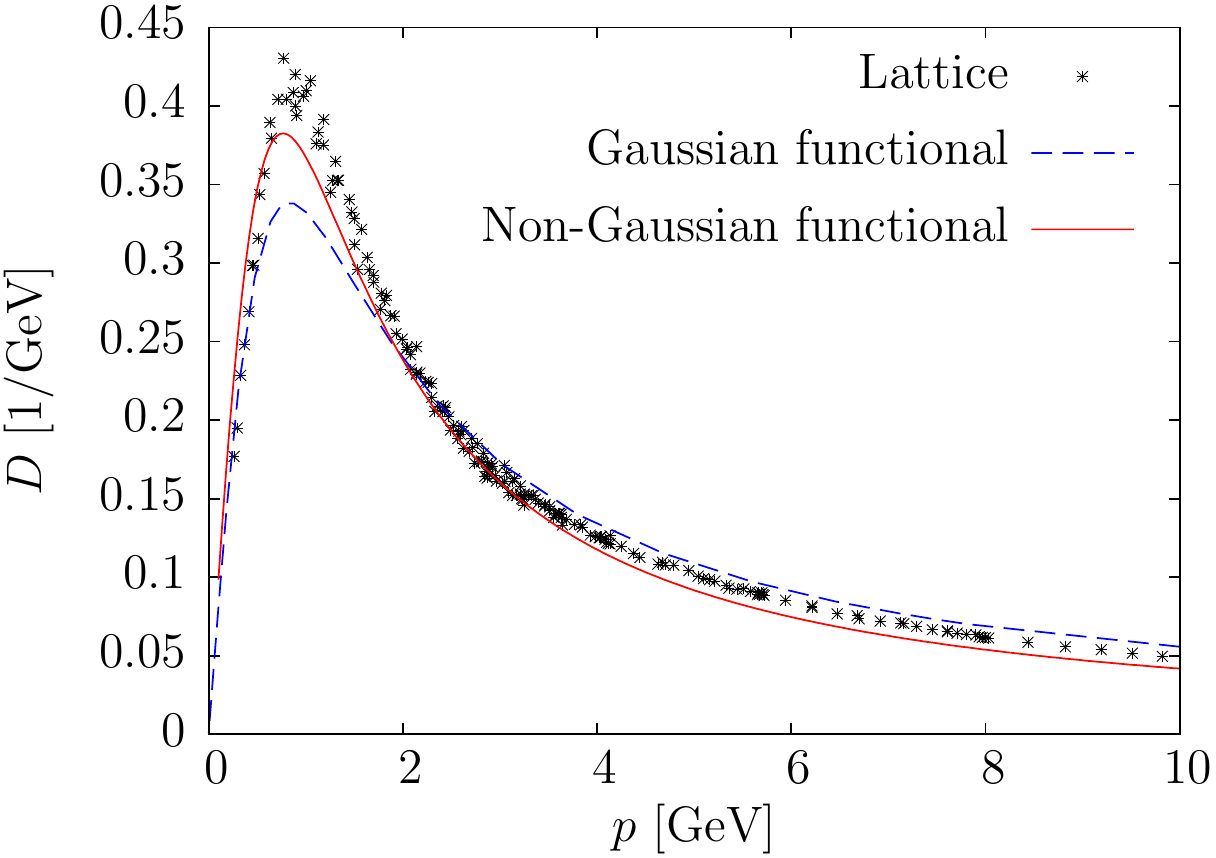}
\caption{The static gluon propagator in Coulomb gauge calculated on the lattice for SU(2) gauge theory (crosses). The dashed and the full curves show the result of the variational 
calculation using, respectively, a Gaussian and non-Gaussian ansatz for the vacuum wave functional.}
\label{fig5}%
\end{figure}%

Let us now compare the results of the variational solution with lattice calculations. Figure \ref{figX} shows the gluon energy and the ghost form factor in $d = 2$ spatial dimension obtained in the variational approach \cite{Feuchter:2007mq} together with the lattice data \cite{Moyaerts_thesis}. The agreement is in general quite satisfactory, in particular, in the IR and  the UV. There are, however, significant deviations in  the mid-momentum regime. A similar picture is obtained in $d = 3$ \cite{BQR2009}. Figure \ref{fig5} shows the static gluon propagator $D = 1/(2 \omega)$ in Coulomb gauge obtained in SU(2) gauge theory in $d = 3$. It is remarkable that the lattice data can be nicely fitted by Gribov's formula \cite{Gribov1978} (see fig.~\ref{fig:gribov:gluonprop})
\beq
\omega(p) = \sqrt{p^2 + \frac{M^4}{p^2}} \, , \label{G13}
\eeq
where $M$ is the so-called Gribov mass. Using a Wilsonian string tension of $\sigma_{\mathrm{W}} = (440 \, \mathrm{MeV})^2$ one finds $M \simeq 880 \, \mathrm{MeV}$. The variational calculations reproduce the infrared behavior of the lattice propagator perfectly and are also in reasonably agreement with the lattice data in the ultraviolet. However, in the mid-momentum regime some strength is missing in the variational calculation. This missing strength is the result of the Gaussian type ansatz for the vacuum wave functional. In Ref.~\cite{CR2010}, the ansatz for the vacuum wave functional was extended to include also cubic and quartic terms of the gauge field in the exponent of the vacuum wave functional,
\begin{subequations}
\begin{align}
\phi[A] &\sim \exp\bigl[-S[A]\bigr] \, , \\
S[A] &= \frac{1}{2} \int A \omega A + \frac{1}{3!} \int \gamma^{(3)} A A A + \frac{1}{4!} \int \gamma^{(4)} A A A A \, ,
\end{align}
\end{subequations}
and one finds the full  curve in fig.~\ref{fig5}, which gives a much better agreement with the lattice data in the mid-momentum regime.

The lattice calculation of the ghost form factor $d(p)$ (\ref{246}) is more involved than that of the gluon propagator since it requires the inversion of the Faddeev--Popov operator $(- \hat{\vD} \cdot \vec{\partial})$, which requires high numerical accuracy for field configurations near the Gribov horizon, where the Faddeev--Popov operator has a very small eigenvalue. It turns out that the lattice results for the ghost form factor depend on how the Coulomb gauge is implemented on the lattice. In principle, this is done by maximizing the gauge fixing functional
\beq
F_t[g] = \sum_{\vx, i} \mathrm{Re} \, \mathrm{tr} U_i^g(t, \vx) \to \max \label{513-f7-G1}
\eeq
with respect to all spatial gauge transformations $g(\vx)$. In eq.~(\ref{513-f7-G1}) the summation is over all spatial links at a fixed time $t$ and the maximization is performed at all lattice times. In the continuum limit the extremum condition $\delta F[g] / \delta g(\vx) = 0$ yields the Coulomb gauge $\vec{\partial} \cdot \vA = 0$.

\begin{figure}
\centering
\begin{subfigure}{6cm}
\includegraphics[width=\textwidth]{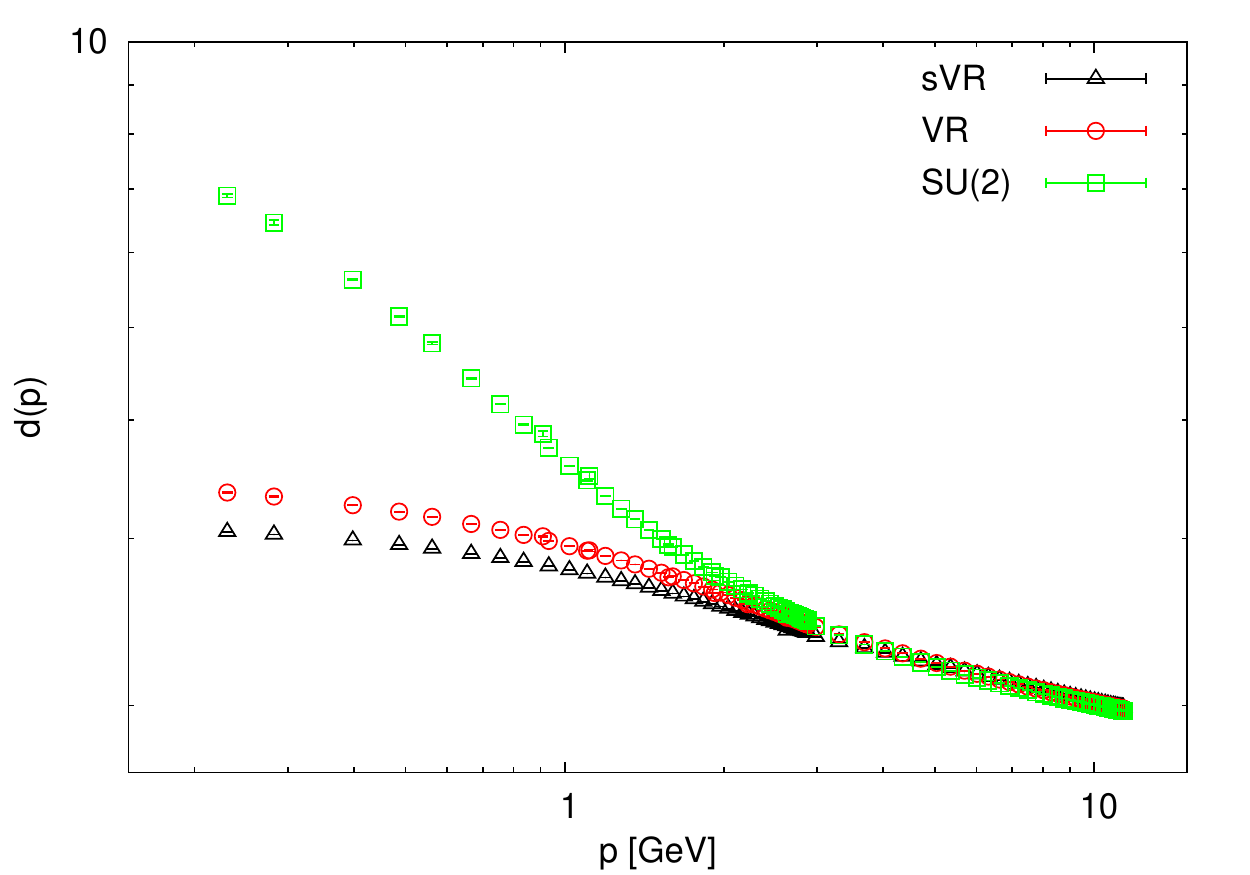}
\caption{}
\end{subfigure}
\quad
\begin{subfigure}{6cm}
\includegraphics[width=\textwidth]{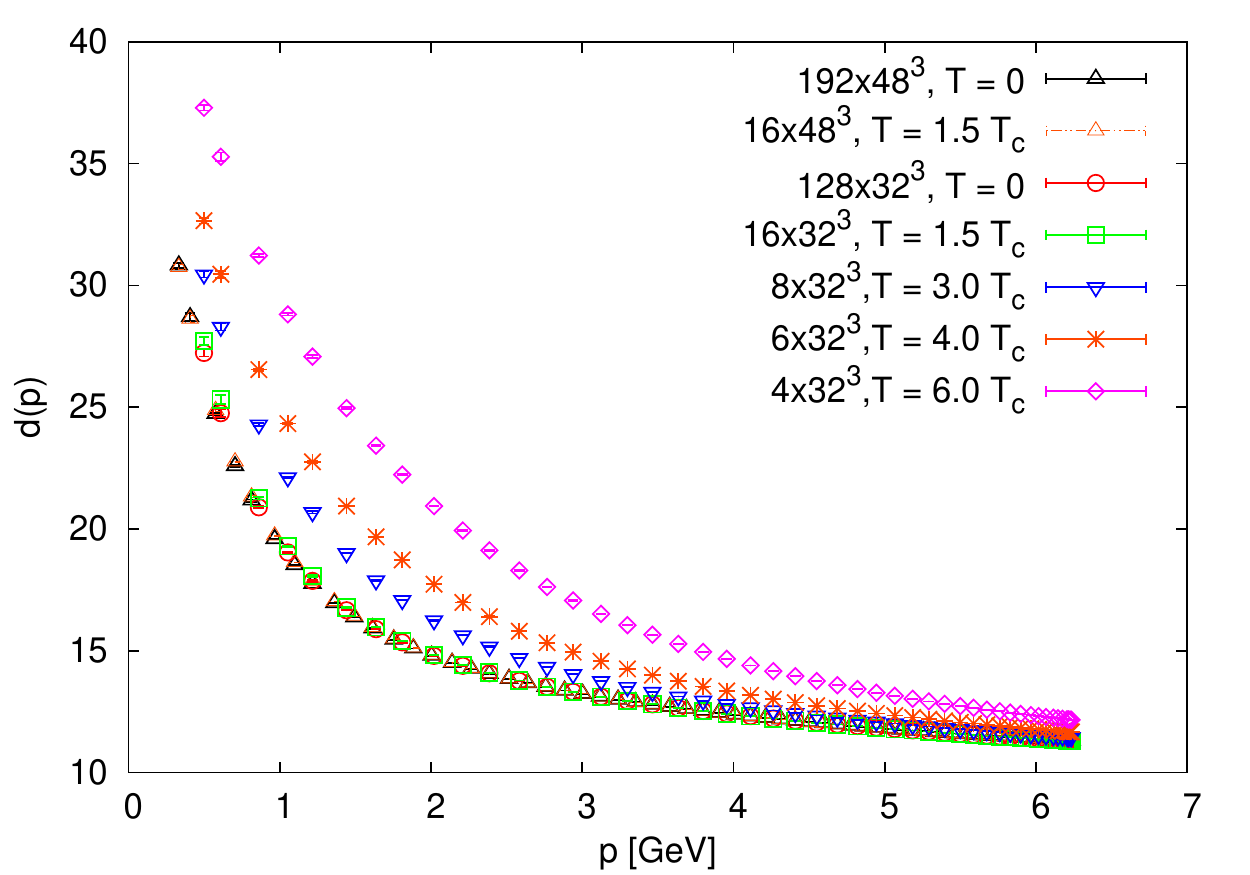}
\caption{}
\end{subfigure}
\caption{(a) The ghost form factor in Coulomb gauge calculated on the lattice in Ref.~\cite{BQRV2015} (green squares). The red circles and black triangles show the results obtained for the ghost form factor when all center vortices or only the spatial center vortices are removed from the ensemble of gauge field configurations, see main text. (b) Ghost form factor calculated on the lattice for different temperatures.}
\label{fig4}%
\end{figure}%

The lattice gauge fixing condition yields a gauge copy which lies within the first Gribov region,
but usually not within the fundamental modular region composed of those copies which 
give the \emph{absolute} maximum of the functional (\ref{513-f7-G1}). Therefore, in practice one repeatedly performs random gauge transformations and selects in the end, i.e.~after Coulomb gauge fixing, the gauge copy which yields the largest maximum of the gauge fixing functional (\ref{513-f7-G1}). This copy is called ``best copy'' since it is assumed that this method yields a gauge copy which is the best representative of the global maximum. Figure \ref{fig4} shows the result for the ghost form factor using the ``best copy'' gauge fixing. The obtained ghost form factor has an IR exponent of $\beta \simeq 0.5$, which is at odds with the sum rule (\ref{314-x2}) given that an IR exponent of $\alpha = 1$ is obtained for the lattice gluon propagator, see eq.~(\ref{G13}). This result is puzzling since the sum rule is considered 
incontrovertible as it is obtained under quite mild assumptions. However, in Ref.~\cite{deForcrand:1994mz} it was shown that for the $U(1)$ lattice gauge theory on $S^2$ the ``best copy'' method does not necessarily provide the best approximation to the fundamental modular region. An alternative lattice gauge fixing method consists in choosing not the ``best'' Gribov copy but that gauge copy which minimizes the lowest eigenvalue of the Faddeev--Popov operator \cite{Sternbeck:2012mf}. This configuration is referred to as the ``lowest'' Gribov copy. As argued in Ref.~\cite{Cooper:2015sza}, the ``lowest copy'' method should yield results closer to the continuum theory. We have used the ``lowest copy''  (\emph{lc}) method to recalculate the ghost and gluon propagator, see Ref.~\cite{Burgio:2016nad}. While the gluon propagator is basically the same as obtained with the ``best copy'' (\emph{bc}) method (see fig.~\ref{fig:gribov:gluonprop}), 
the ghost form factor gets further enhanced in the IR as the number of 
gauge-fixing attempts\footnote{To find the absolute extremum in the bc and lc approach, we 
have repeated the gauge-fixing procedure a large number $N_r = 10,\ldots,10000$ of times, starting 
each time from a different random gauge transformation of the original configuration. In general, 
the number $N_r$ of gauge-fixing trials is indicative of the number of Gribov copies 
included, even though the exact relation is complicated and non-linear \cite{Burgio:2016nad}.}
increases (see fig.~\ref{fig:gribov:ghost:lc:n24t24_allcopies}).
Although we did not find a strict saturation for a sufficiently large number of gauge-fixing 
attempts, the IR exponent of the ghost form factor is compatible with the continuum result of 
$\beta \simeq 1$ (see fig.~\ref{fig:gribov:ghost:n24t24lcbc}), in agreement with the sum rule (\ref{314-x2}).

\begin{figure}[hptb]
\center
\includegraphics[width=0.49\columnwidth]{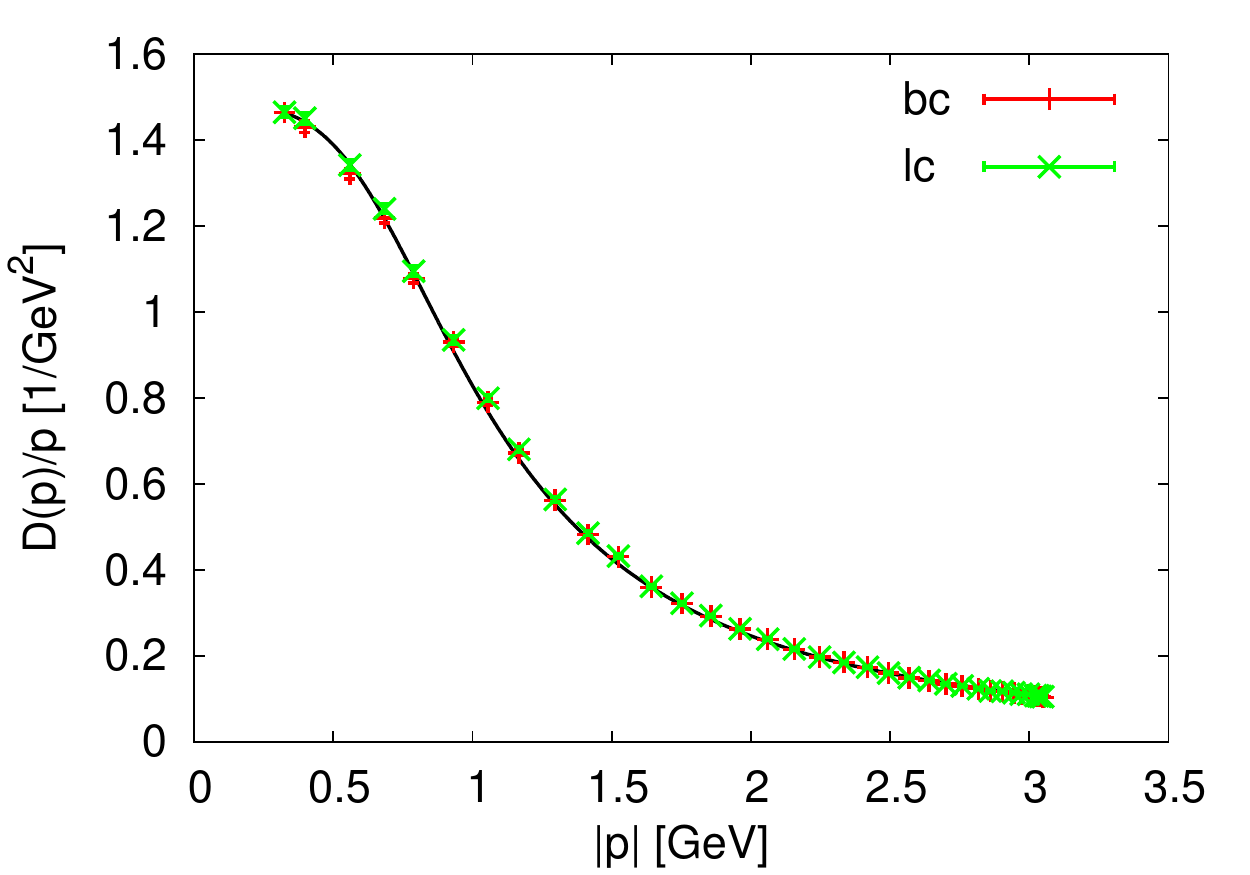}
\caption{The gluon propagator with the bc and the lc-approach from $1000$ gauge-fixing attempts. 
The solid line is a fit to the Gribov formula (\ref{G13}). The choice of Gribov copies apparently 
makes no visible difference.}
\label{fig:gribov:gluonprop}
\end{figure}

\begin{figure}[phtb]
\centering
\begin{subfigure}{0.49\textwidth}
\includegraphics[width=\textwidth]{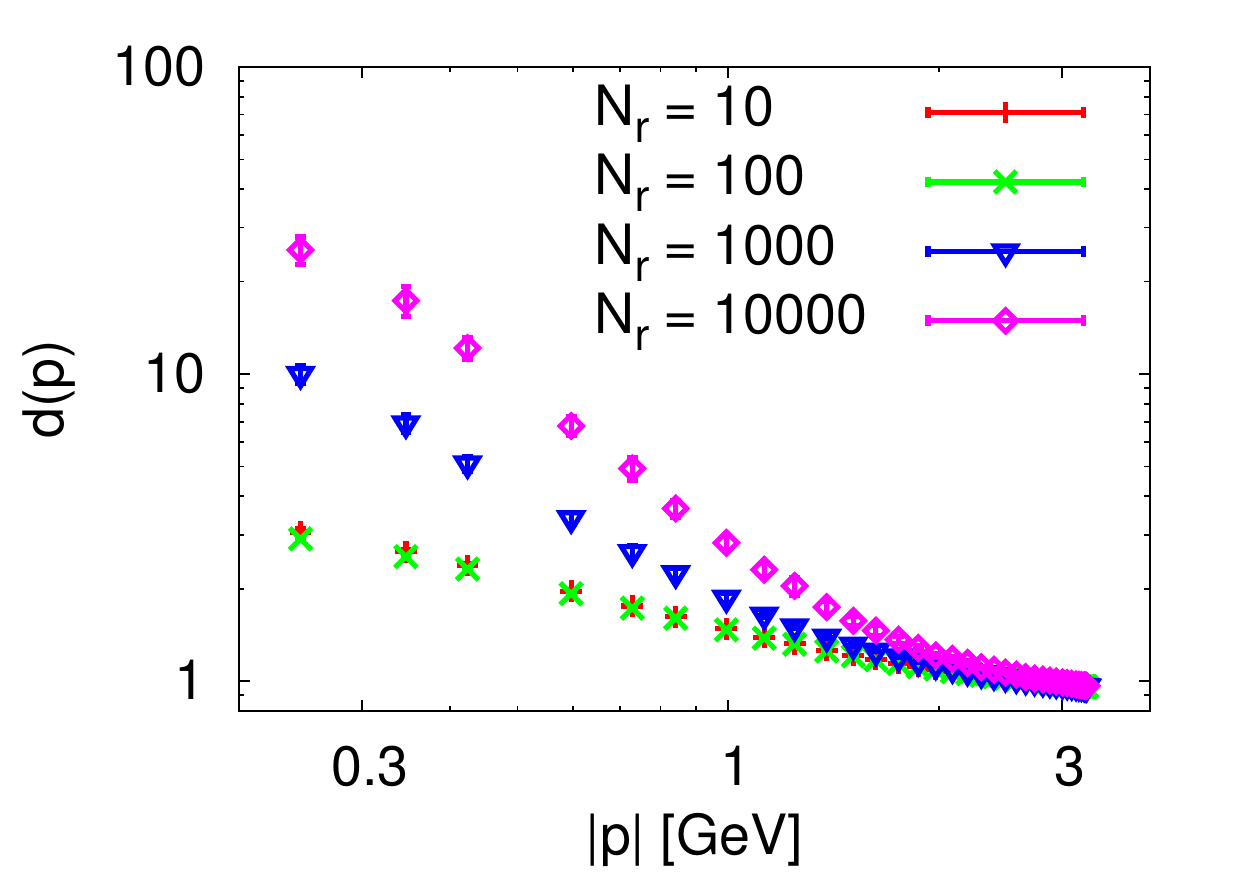}
\caption{}
\end{subfigure}
\hfill
\begin{subfigure}{0.49\textwidth}
\includegraphics[width=\textwidth]{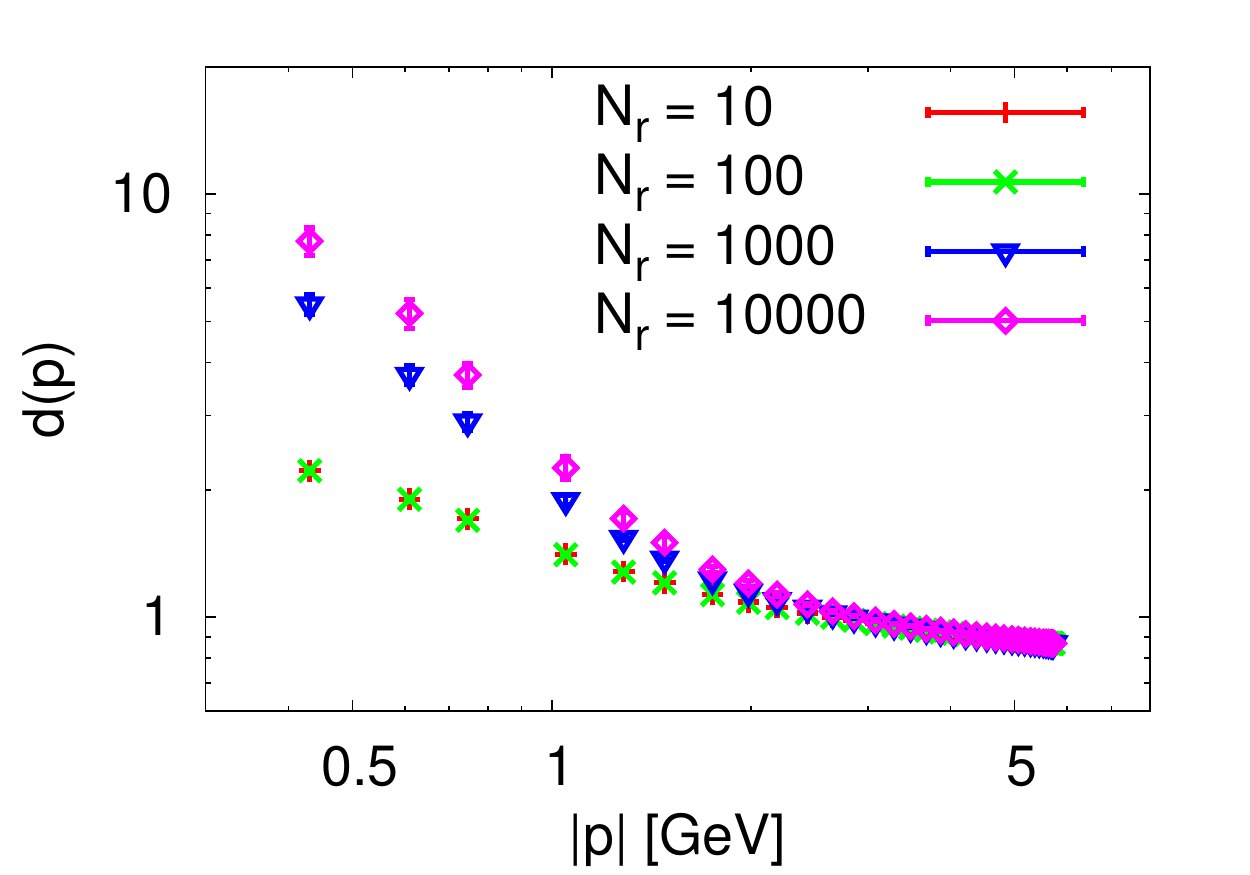}
\caption{}
\end{subfigure}
\caption{The ghost form factor after gauge fixing to the lowest-eigenvalue copy with increasing number of trials from $10$ to $10000$ on $24^4$ lattices at (a) $\beta = 2.2$ and (b) $\beta = 2.4$.}
\label{fig:gribov:ghost:lc:n24t24_allcopies}
\end{figure}

\begin{figure}[phtb]
\centering
\includegraphics[width=0.49\columnwidth]{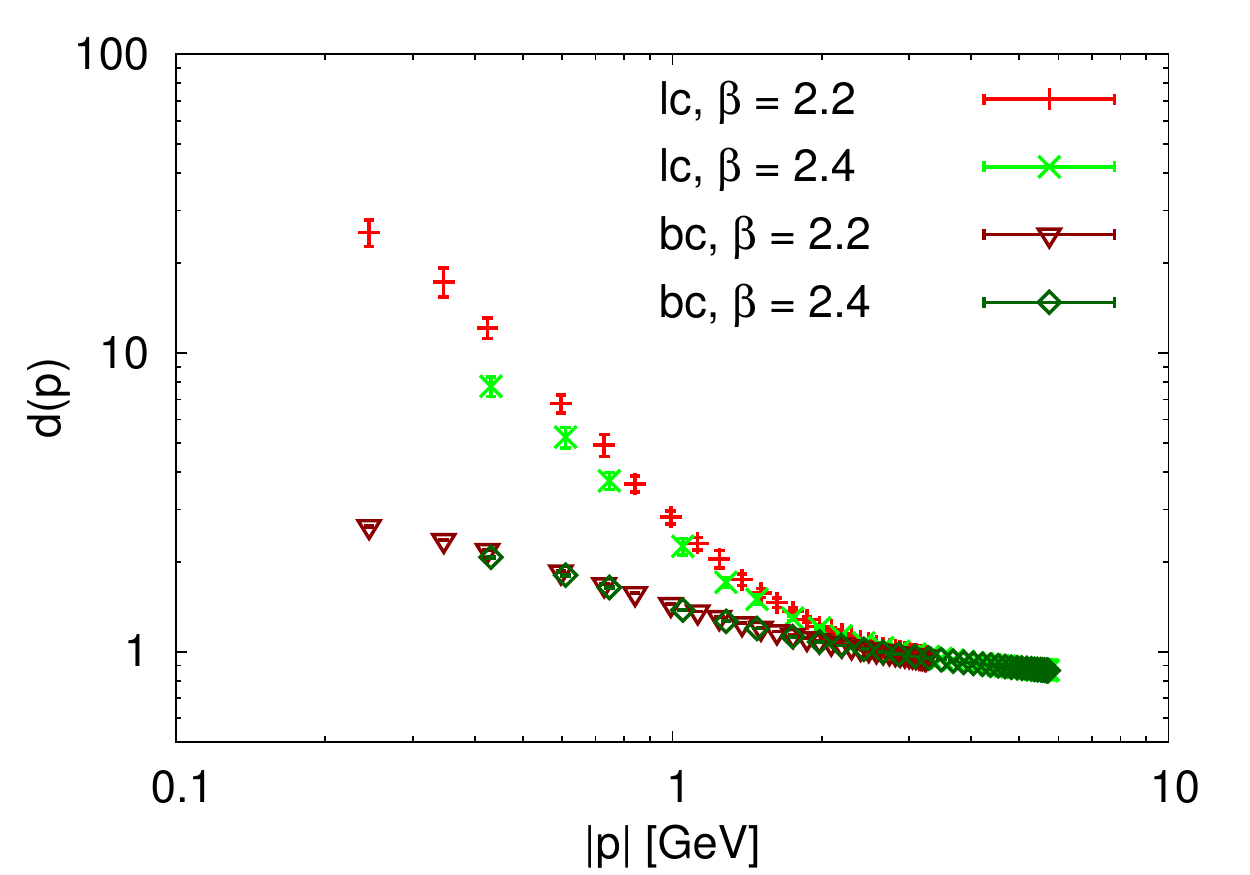}
\caption{The ghost form factor after $10000$ copies of bc and lc-strategy.}
\label{fig:gribov:ghost:n24t24lcbc}
\end{figure}

\section{The Coulomb string tension}

\begin{figure}
\centering
\begin{subfigure}{0.45\textwidth}
\includegraphics[width=\textwidth,clip]{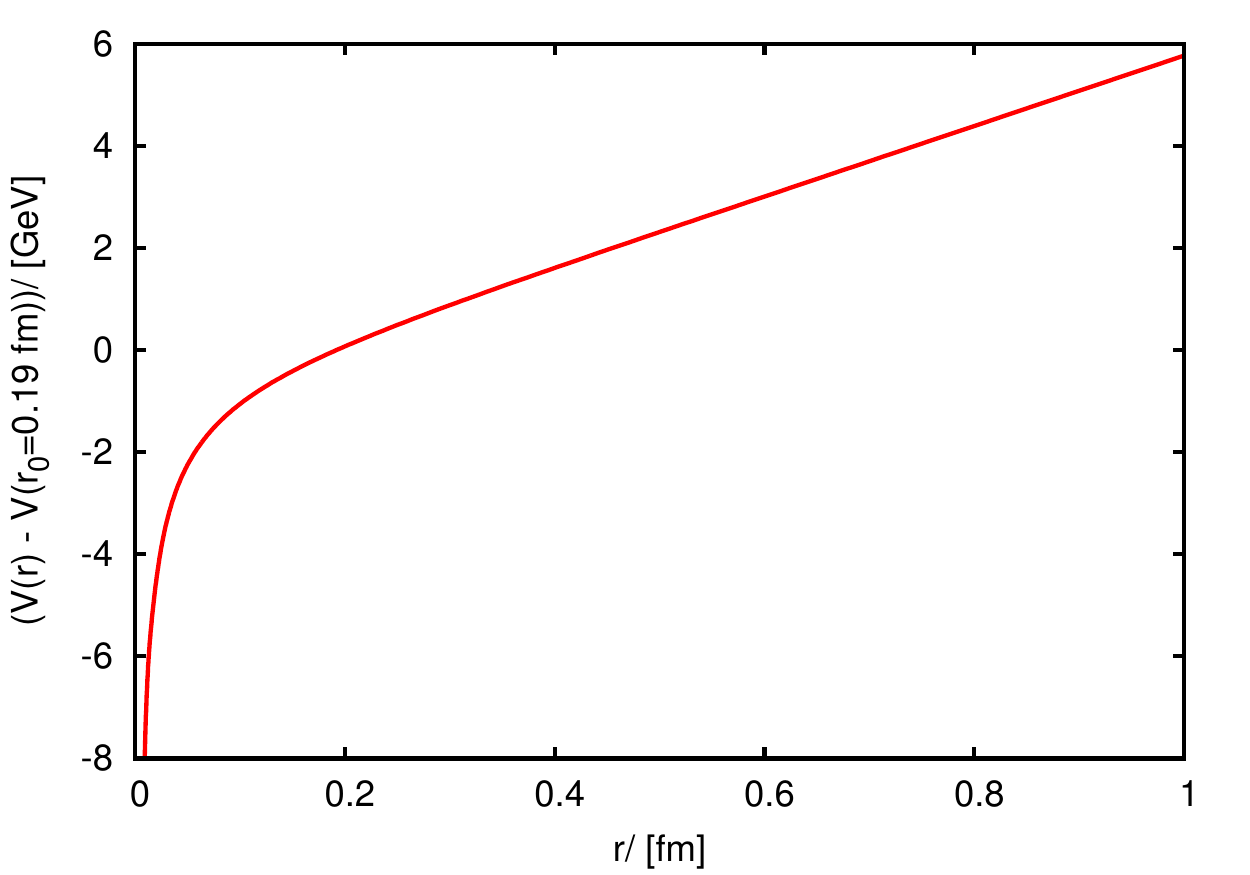}
\caption{}
\end{subfigure}
\quad
\begin{subfigure}{0.45\textwidth}
\includegraphics[width=\textwidth,clip]{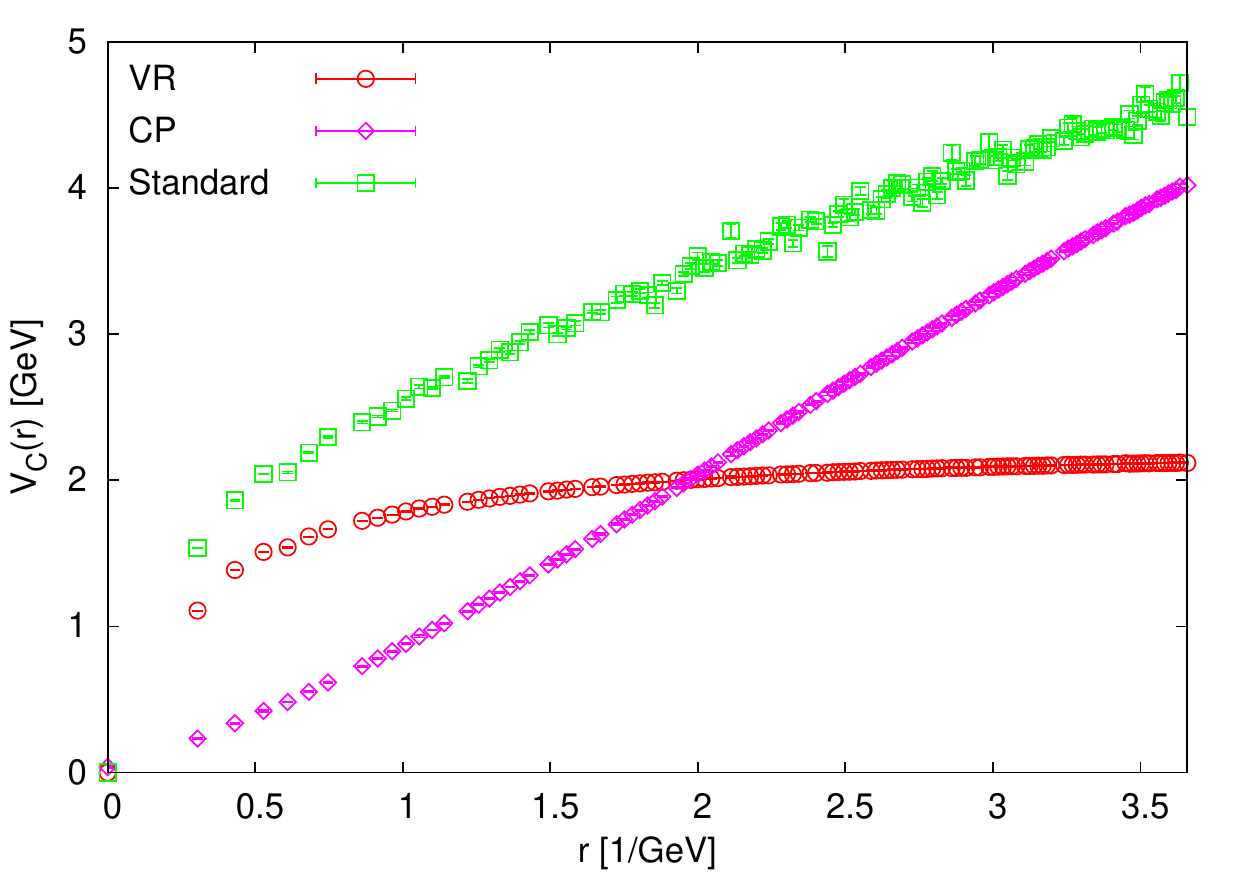}
\caption{}
\end{subfigure}
\caption{(a) Non-Abelian Coulomb potential (\ref{Gx}) obtained within the variational approach \cite{ERS2007}. (b) Standard non-Abelian Coulomb potential (green boxes) compared to the potential obtained after vortex removal (red circles) and center projection (violet diamonds) \cite{BQRV2015}.}
\label{fig2ab}%
\end{figure}

The Coulomb term $H_{\mathrm{C}}$ (\ref{2}) plays an important role in the Gribov--Zwanziger confinement scenario. Its Yang--Mills vacuum expectation value
\beq
V_{\mathrm{C}} = g^2 \bigl\langle (-\hat{\vD} \cdot \vec{\partial})^{-1} (-\vec{\partial}^2) (-\hat{\vD} \cdot \vec{\partial})^{-1} \bigr\rangle \label{Gx}
\eeq
provides an upper bound for the potential between static point-like color charges and is referred to as (non-Abelian) Coulomb potential. The Coulomb potential found within the variational approach \cite{Feuchter2004, Feuchter2005, ERS2007} is shown in fig.~\ref{fig2ab} (a). At small distances it behaves like an ordinary Coulomb potential, $V_{\mathrm{C}}(r) \sim 1/r$, and increases linearly at large distances with a coefficient given by the so-called Coulomb string tension $\sigma_{\mathrm{C}}$. It was shown in Ref.~\cite{Zwanziger2003} that this quantity is an upper bound to the Wilsonian string tension $\sigma_{\mathrm{W}}$. On the lattice one finds $\sigma_{\mathrm{C}} / \sigma_{\mathrm W} \approx 2 \ldots 4$ \cite{BQRV2015, GOZ2004, Voigt2008}. Due to the constraint $\sigma_{\mathrm{C}} \geq \sigma_{\mathrm{W}}$ in the Gribov--Zwanziger confinement scenario a necessary condition for confinement is that the non-Abelian Coulomb potential (\ref{Gx}) rises linearly at large distances.

One may now ask, what field configurations induce the horizon condition, $d^{- 1}(0) = 0$, and the linearly rising Coulomb potential $V_{\mathrm{C}}$ (\ref{Gx}) and thus confinement? 
Given the relation of Gribov's confinement scenario to the dual superconductor, we expect magnetic monopoles to play a substantial role. Lattice calculations carried out in the so-called indirect maximum center gauge, which contains the maximum Abelian gauge in an intermediate step, show that magnetic monopoles are tied to center vortices \cite{Greensite2011}. This can be also understood in the continuum \cite{Reinhardt:2001kf}. Center vortices are string-like gauge field configurations 
in $D=3$ or world surfaces in $D=4$, for which the Wilson loop equals a non-trivial center element of the gauge group, provided the loop has non-trivial linking with the center vortices.\footnote{By the Bianchi identity center vortices form closed loops in $D = 3$ and closed surfaces in $D = 4$.}

Lattice calculations provide strong evidence that confinement is due to center vortices. Indeed, when the center vortex content of the gauge field configurations is removed, one finds that the Wilsonian string tension and thus confinement disappears \cite{deForcrand1999}.

On the lattice center vortices can be detected as follows \cite{DelDebbio1998}: One first brings the gauge field configurations into the so-called maximal center gauge
\beq
\sum_{x, \mu} \bigl\vert \tr \, U_{\mu}^2(x)\bigr\vert \to \max \, , \label{9}
\eeq
which rotates a link as close as possible to a center element, e.g.~$Z_\mu(x) = \pm 1 \in Z (2)$ for the gauge group SU(2). Subsequently, one performs a so-called center projection
\beq
U_\mu(x) \to Z_\mu(x) \label{10}
\eeq
which replaces each link by its nearest center element. One is then left with $Z(2)$ links, which form closed center vortices, the only non-trivial field configurations in a $Z(2)$ theory. When a center vortex pierces a Wilson loop it contributes a non-trivial center element to the latter. It was shown in Ref.~\cite{Langfeld1997} that the center vortices obtained in this way are physical objects in the sense that they show the proper scaling behavior, i.e.~their area density survives the continuum limit. This property distinguishes the center vortices found after center projection in the maximal center gauge from other gauges like e.g. the Laplacian center gauge \cite{Greensite2011}.

The center vortex content of a gauge field configuration can be removed \cite{deForcrand1999} by multiplying the original link variable $U_{\mu}(x)$ by its center projection $Z_{\mu}(x)$,
\beq
U_{\mu}(x) \to U_{\mu}(x) \cdot Z_{\mu} (x) \, . \label{11}
\eeq
Figure \ref{fig4} shows the ghost form factor obtained on the lattice when the center vortices are removed from the ensemble of gauge field configurations as described above \cite{BQRV2015}. The ghost form factor becomes infrared flat and the horizon condition is lost. This shows that center vortices induce the horizon condition which is the cornerstone of Gribov's confinement scenario. This also shows that Gribov's confinement scenario is tied to the center vortex picture of confinement. This is in accord with the observation that center vortices and magnetic monopoles are located on the Gribov horizon of Coulomb gauge \cite{GOZ2004}.

When center vortices are removed as described above the static color potential extracted from a Wilson loop loses its linearly rising part, i.e.~the Wilsonian string tension $\sigma_{\mathrm{W}}$ disappears after center vortex removal. Since $\sigma_{\mathrm{C}} \geq \sigma_{\mathrm{W}}$ this does not necessarily imply that elimination of center vortices also removes the Coulomb string tension. In Ref.~\cite{BQRV2015} the non-Abelian Coulomb potential was calculated after center projection and center vortex removal. Removing the center vortices also eliminates the Coulomb string tension while center vortex projection keeps only the linearly rising part of the non-Abelian Coulomb potential, see fig.~\ref{fig2ab} (b). This result is perhaps not so surprising since center vortices live on the Gribov horizon,\footnote{More precisely on the common boundary between the Gribov horizon and the fundamental modular region \cite{GOZ2004}.} which represents the domain of the infrared dominant field configurations in the Gribov--Zwanziger confinement scenario.

\begin{figure}
\centering
\begin{subfigure}{0.45\textwidth}
\includegraphics[width=\textwidth,clip]{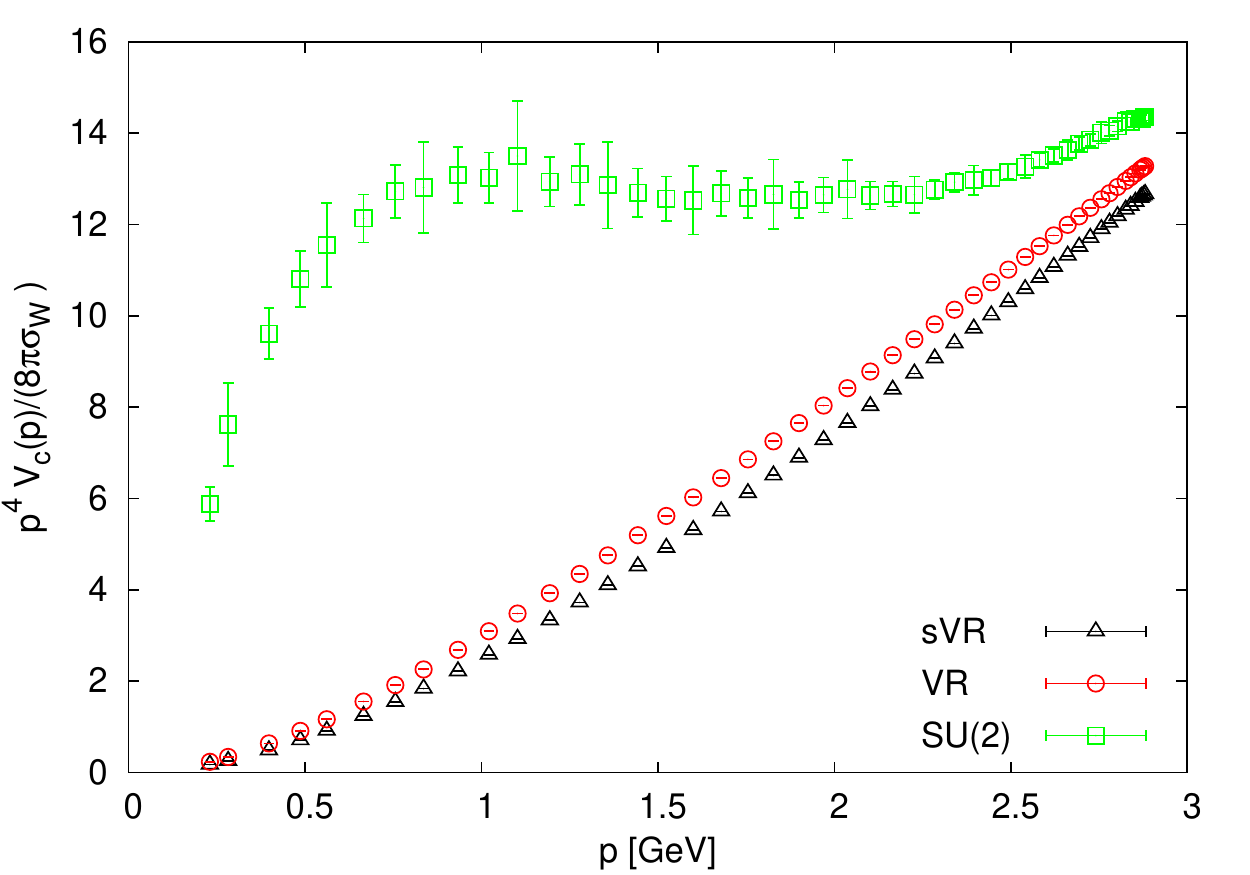}
\caption{}
\end{subfigure}
\quad
\begin{subfigure}{0.45\textwidth}
\includegraphics[width=\textwidth,clip]{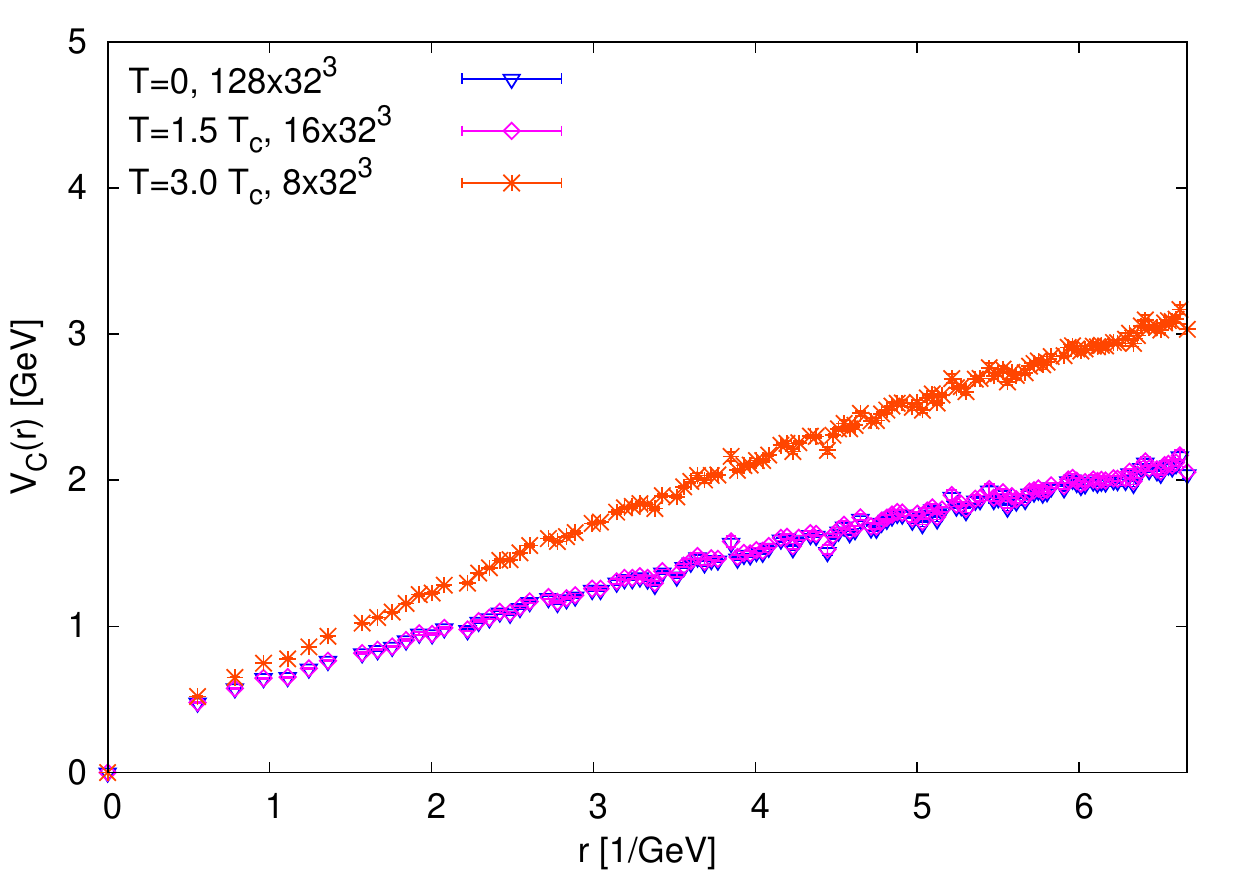}
\caption{}
\end{subfigure}
\caption{(a) Non-Abelian Coulomb potential in momentum space (green boxes) compared to the result obtained after removing just the spatial (black triangles) or all center vortices (red circles). (b) Non-Abelian Coulomb potential for different temperatures ($T_{\mathrm{c}}$ denotes the critical temperature).}
\label{fig3ab}%
\end{figure}

At finite temperature different Wilsonian string tensions are measured from temporal and spatial Wilson loops referred to as temporal and spatial string tension, respectively. Above the deconfinement phase transition these two Wilsonian string tensions decouple. While the spatial string tension increases above the critical temperature, the temporal string tension disappears. On the lattice it is not difficult to see that in the center projected $Z(2)$ theory the temporal and spatial Wilsonian string tension, i.e.~the area law in the temporal and spatial Wilson loop, are produced by temporal and spatial center vortices, respectively. The latter are formed exclusively by spatial center-valued 
links \footnote{In $D=3$, spatial center vortices are closed lines formed by a "stack" of spatial plaquettes with non-trivial value after center projection. Geometrically, they extend in the time direction (on the dual lattice) and can thus link with spatial Wilson loops. The terminology in 
$D=4$ is similar, i.e.~spatial center vortices are hyper-surfaces on the dual lattice which are 
composed of spatial plaquettes on the original lattice, which are non-trivial after center projection. Geometrically, such spatial vortices extend in one space and one time direction and may hence link with spatial Wilson loops.}
\beq
U_i (x) \to Z_i (x) \, , \label{12}
\eeq
which will be referred to as spatial center projection in the following. Analogously multiplying the spatial link by its nearest center projected $Z(2)$ element,
\beq
U_i(x) \to U_i(x) \cdot Z_i(x) \, , \label{13}
\eeq
removes all spatial center vortices and thus the spatial string tension while the temporal links are unaffected. Therefore, the temporal string tension, which can be calculated from the correlator of Polyakov loops and hence from temporal links exclusively, will not be affected by the spatial center vortex removal. Figure \ref{fig3ab} (a) shows the quantity $p^4 V_{\mathrm{C}}(p)$ whose infrared limit gives the Coulomb string tension, $\lim_{p \to 0} p^4 V_{\mathrm{C}}(p) = 8 \pi \sigma_{\mathrm{C}}$. As one observes, the Coulomb string tension disappears already when only the spatial center vortices are removed. This clearly shows that the Coulomb string tension is related to the spatial string tension and not to the temporal one. This explains also the finite-temperature behavior of the Coulomb string tension, which increases with the temperature above the deconfinement phase transition just like the spatial string tension, see fig.~\ref{fig3ab} (b).

A necessary condition for the Gribov--Zwanziger confinement scenario to be realized is that the ghost form factor is infrared divergent, which is indeed found in the variational approach and also on the lattice, see fig.~\ref{fig4} (b). However, the infrared divergence disappears when one removes the center vortices or the spatial center vortices only as can be seen in fig.~\ref{fig4} (a). Also, the spatial center vortex projection produces the same ghost form factor as full center projection. This also explains why the infrared divergence of the ghost form factor does not disappear above the deconfinement phase transition. Thus both features of the Gribov--Zwanziger confinement scenario, the infrared divergent ghost form factor and the linearly rising Coulomb potential, are caused by spatial center vortices and are thus tied to the spatial string tension, which increases above 
the deconfinement transition.

\section{Variational approach to the quark sector}

The variational approach to Yang--Mills theory in Coulomb gauge presented in section \ref{sectII} has been extended to full QCD in Refs.~\cite{Pak2013, QCDT0, QCDT0Rev}. The Hamiltonian of full QCD in Coulomb gauge is given by
\beq
H_{\mathrm{QCD}} = H_{\mathrm{T}} + H_{\mathrm{Q}} + H_{\mathrm{C}} \, , \label{14}
\eeq
where $H_{\mathrm{T}}$ is the Hamiltonian of the transversal gluon degrees of freedom (\ref{G1}), $H_{\mathrm{C}}$ is the Coulomb interaction (\ref{2}) and
\beq
H_{\mathrm{Q}} = \int \dd^3 x \, \psi^\dagger(\vx) \Bigl[\vec{\alpha} \cdot \bigl(-\ii \nabla + g t^a \vA^a(\vx)\bigr) + \beta m_0\Bigr] \psi(\vx) \label{15}
\eeq
is the Hamiltonian of the quarks coupling to the transversal gluon field. Here, $\vec{\alpha}$, $\beta$ are the usual Dirac matrices, $t^a$ denotes the generator of the color group in the fundamental representation and $m_0$ is the bare current quark mass (of electroweak origin) which will be neglected in the following. Furthermore, when the quarks are included, the matter charge density in the Coulomb Hamiltonian $H_{\mathrm{C}}$ (\ref{3}) is given by
\beq
\rho_m^a(\vx) = \psi^\dagger(\vx) t^a \psi(\vx) \, . \label{16}
\eeq
In Refs.~\cite{QCDT0, QCDT0Rev}, the quark sector of QCD has been treated within the variational approach using the following ansatz for the QCD wave functional
\beq
\vert \phi[A] \rangle = \cN \frac{1}{\sqrt{I[A]}} \phi_{\mathrm{YM}}[A] \, \vert \phi_{\mathrm{Q}}[A] \rangle \label{18}
\eeq
where $\phi_{\mathrm{YM}}$ is the Yang--Mills vacuum functional (\ref{G6}) and
\beq
\vert \phi_{\mathrm{Q}}[A] \rangle = \exp\left[-\int \dd^3 x \int \dd^3 y \, \psi_+^{\dagger}(\vx) K(\vx, \vy) \psi_-(\vy)\right] \vert 0 \rangle \, , \label{17}
\eeq
with
\beq
K(\vx, \vy) = \beta S(\vx, \vy) + g \int \dd^3 z \, \bigl[V(\vx, \vy; \vz) + \beta W(\vx, \vy; \vz)\bigr] \vec{\alpha} \cdot \vA^a(\vz) t^a \label{ansx}
\eeq
is the quark wave functional. Here, $S$, $V$ and $W$ are variational kernels. Furthermore, $\vert 0 \rangle$ is the Fock vacuum of the quarks which represents the bare Dirac sea. Finally,
\beq
\label{801-15}
I[A] = \langle \phi_{\mathrm{Q}}[A] \vert \phi_{\mathrm{Q}}[A] \rangle = \det(\mathrm{id} + K^\dagger K)
\eeq
is the quark determinant. The ansatz (\ref{18}) treats the quark determinant $I[A]$ and the Faddeev--Popov determinant (\ref{203-1}) on equal footing.

The ansatz (\ref{17}) reduces for $W = 0$ to the quark wave functional used in Ref.~\cite{Pak2013} while for $V = W = 0$ it becomes the BCS-type wave functional considered in Refs.~\cite{FM1982, Adler1984, AA1988}. With the wave functional (\ref{18}) the expectation value of the QCD Hamiltonian was calculated up to two loops. Variation with respect to the two kernels $V$ and $W$, which describe the coupling of the quarks to the transversal gluons, gives two equations, which can be solved explicitly in terms of the scalar kernel $S$ and the gluon energy $\omega$ yielding
\begin{align}
V(\vp, \vq) &= \frac{1 + S(p) S(q)}{p P(p) \Bigl(1 - S^2(p) + 2 S(p) S(q)\Bigr) + q P(q) \Bigl(1 - S^2(q) + 2 S(p) S(q)\Bigr) + \omega(|\vp + \vq|)} \, , \label{19} \\
W(\vp, \vq) &= \frac{S(p) + S(q)}{p P(p) \Bigl(1 - S^2(p) - 2 S(p) S(q)\Bigr) + q P(q) \Bigl(1 - S^2(q) - 2 S(p) S(q)\Bigr) + \omega(|\vp + \vq|)} \label{20}
\end{align}
where we have defined the quantity
\beq
P(p) = \frac{1}{1 + S^2(p)} \, . \label{23}
\eeq
The variational equation for the scalar kernel $S$, referred to as \textit{gap equation}, is highly non-local and can only be solved numerically. However, one can show analytically that all UV divergences in this equation cancel: the UV-divergent contributions to $S(k)$ induced by the kernels $V$ and $W$ are given, respectively, by
\begin{align}
&\frac{C_{\mathrm{F}}}{16 \pi^2} g^2 S(k) \left[-2 \Lambda + k \ln \frac{\Lambda}{\mu} \left(-\frac{2}{3} + 4 P(k)\right)\right] , \label{21} \\
&\frac{C_{\mathrm{F}}}{16 \pi^2} g^2 S(k) \left[2 \Lambda + k \ln \frac{\Lambda}{\mu} \left(\frac{10}{3} - 4 P(k)\right)\right] . \label{22}
\end{align}
Here, $C_{\mathrm{F}} = (N_{\mathrm{C}}^2 - 1) / 2 N_{\mathrm{C}}$ is the quadratic Casimir, $\Lambda$ is the UV cutoff and $\mu$ is an arbitrary momentum scale. In the sum of the two terms given by eq.~(\ref{21}) and (\ref{22}), the linear UV divergences obviously cancel. Furthermore, the sum of the logarithmic UV divergences of these two terms cancels against the asymptotic contribution to the gap equation induced by the Coulomb kernel,
\beq
-\frac{C_{\mathrm{F}}}{6 \pi^2} g^2 k S(k) \ln \frac{\Lambda}{\mu} \, . \label{24}
\eeq

\begin{figure}
\centering
\begin{subfigure}{0.45\textwidth}
\includegraphics[width=\textwidth,clip]{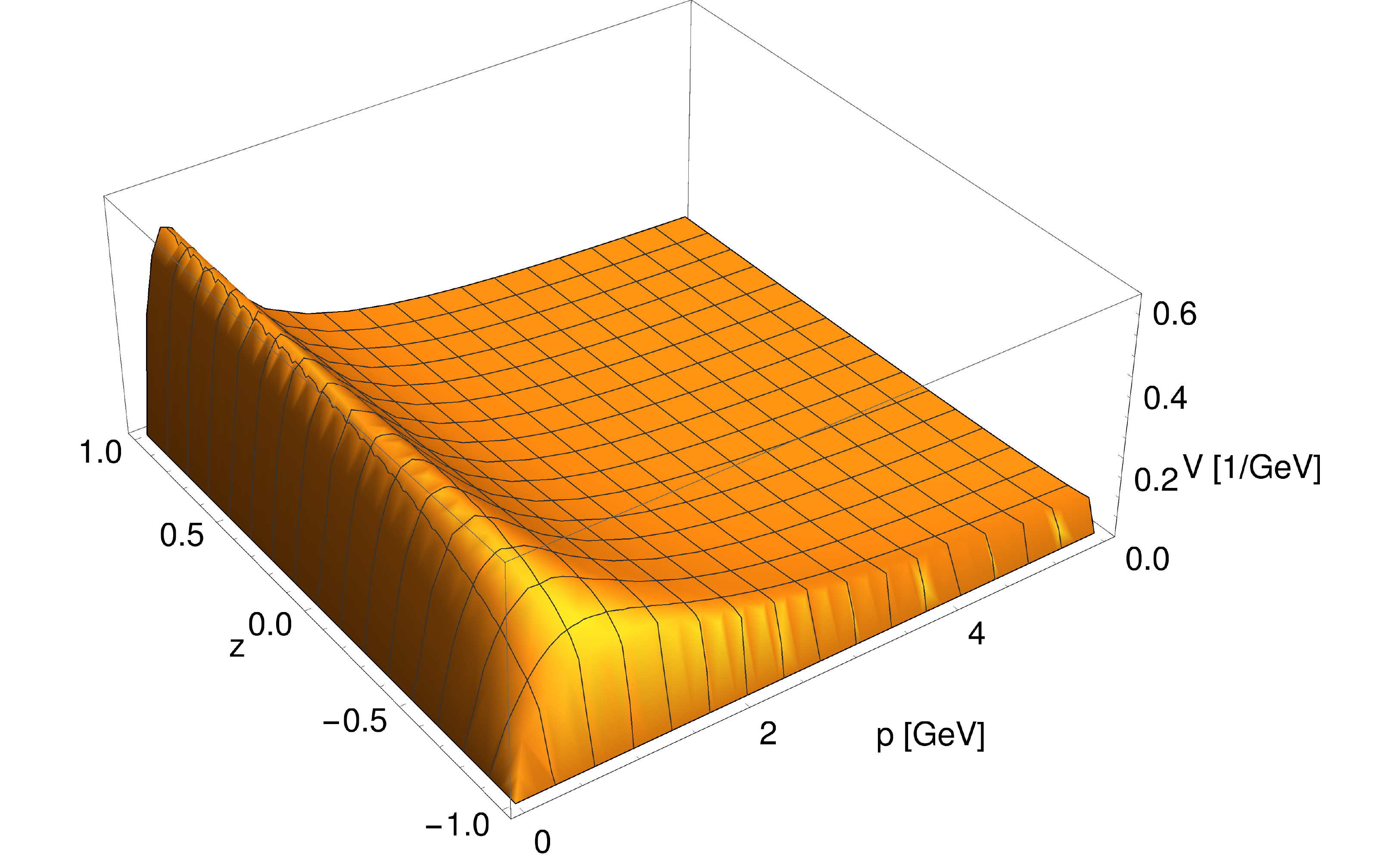}
\caption{}
\end{subfigure}
\quad
\begin{subfigure}{0.45\textwidth}
\includegraphics[width=\textwidth,clip]{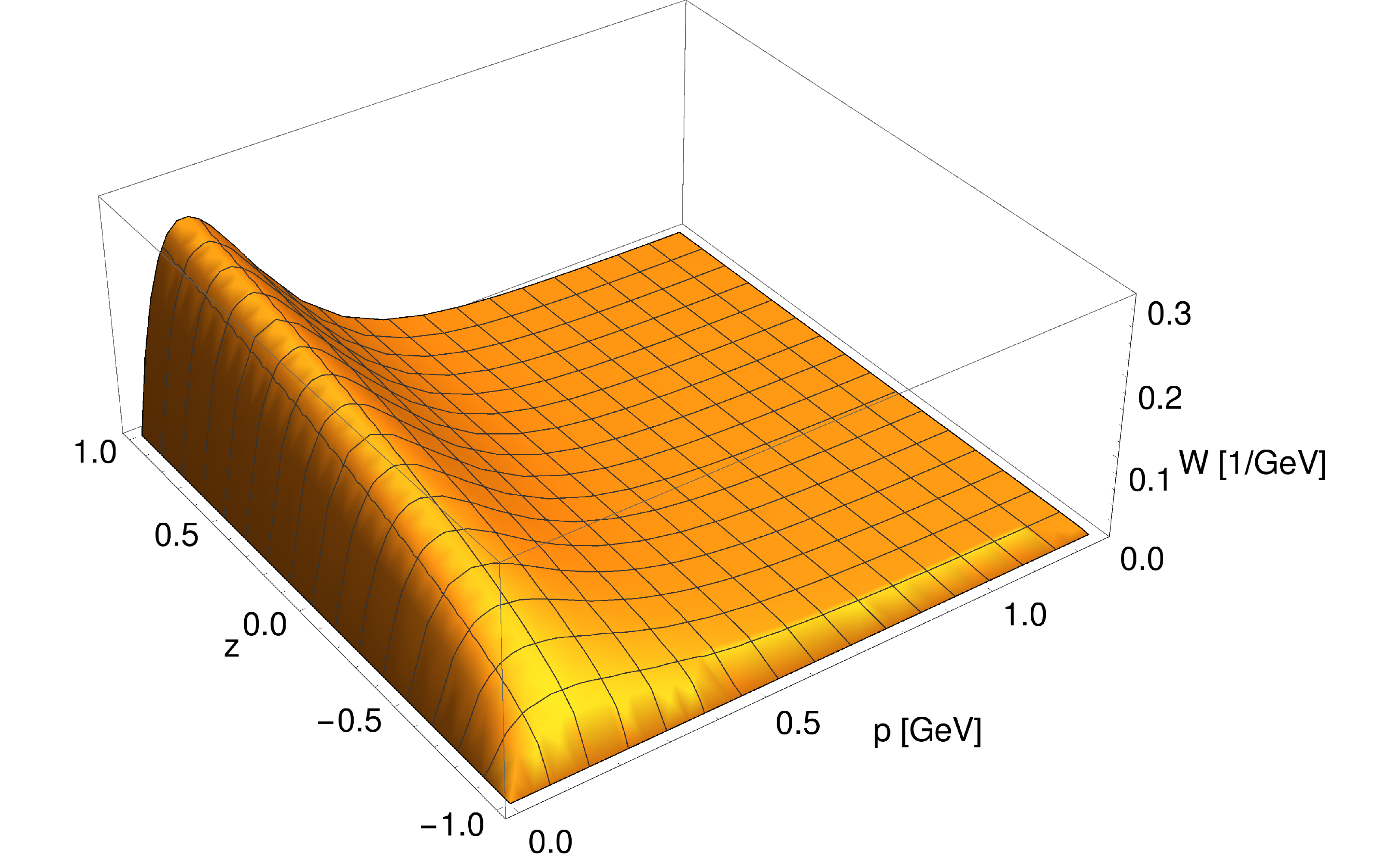}
\caption{}
\end{subfigure}
\caption{The vector kernel (a) $V(\vp, \vq)$ (\ref{19}) and (b) $W(\vp, \vq)$ (\ref{20}) obtained from the solution of the quark gap equation for $g \simeq 2.1$ as function of $p = q$ and $z = \cos\sphericalangle(\vp,\vq)$ \cite{QCDT0Rev}.}
\label{fig4ab}%
\end{figure}%

\begin{figure}
\centering
\begin{subfigure}{0.45\textwidth}
\includegraphics[width=\textwidth,clip]{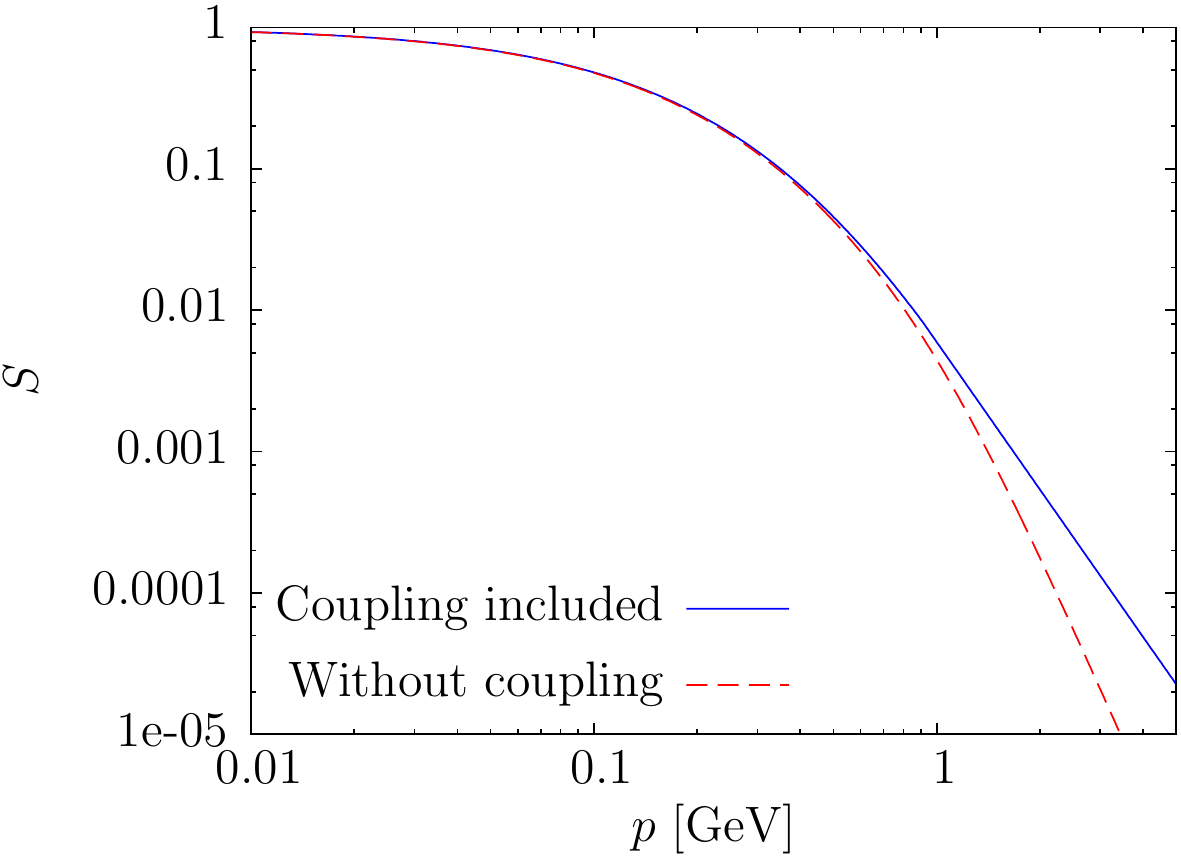}
\caption{}
\end{subfigure}
\quad
\begin{subfigure}{0.45\textwidth}
\includegraphics[width=\textwidth,clip]{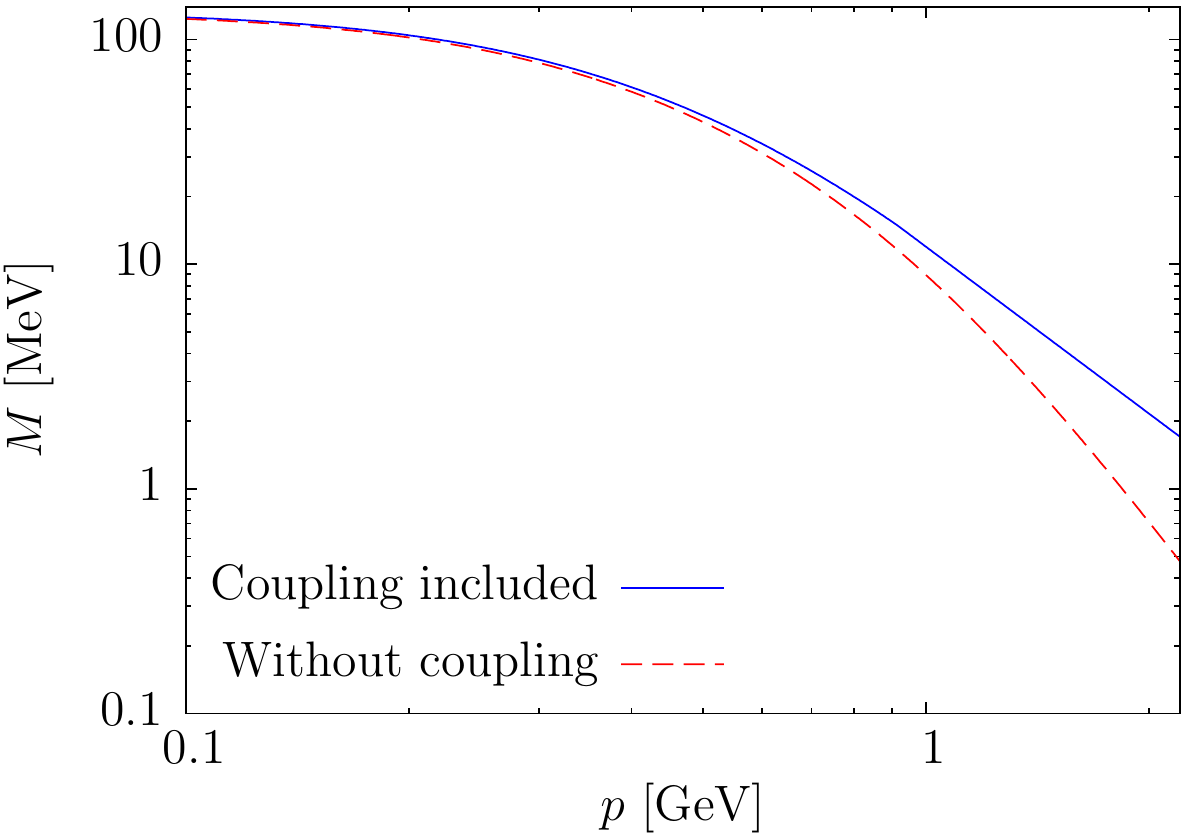}
\caption{}
\end{subfigure}
\caption{(a) Scalar form factor and (b) mass function obtained from the (quenched) solution of the quark gap equation. Results are presented for $g \simeq 2.1$ (full curve) and $g = 0$ (dashed curve).}
\label{fig5ab}%
\end{figure}%

Due to the exact cancellation of all UV divergences no renormalization of the gap equation is required. This is certainly a big advantage of the present ansatz (\ref{17}) for the quark wave functional. Using the gluon propagator $\sim 1/\omega$ obtained in the Yang--Mills sector as input, the quark gap equation can be solved within a quenched calculation. In this approach, the coupling constant $g$ is determined by fixing the chiral quark condensate to its phenomenological value \cite{QCDT0Rev}. Figure \ref{fig4ab} shows the vector kernels $V(\vp, \vq)$, $W(\vp, \vq)$ obtained in this way, as function of the modulus $p = q$ of the ingoing quark momenta and the cosine of the angle between them, $z = \cos\sphericalangle(\vp,\vq)$. These kernels are peaked in the mid-momentum regime. Furthermore, the vector kernel $V$ is about a factor of two larger than the kernel $W$. Figure \ref{fig5ab} shows the scalar kernel $S(p)$ and the mass function
\beq
M(p) = \frac{2 p S(p)}{1 - S^2(p)} \label{25}
\eeq
on a logarithmic scale. For sake of comparison we also quote the curves obtained when the coupling to the transversal gluons is neglected. More precisely, this corresponds to putting $g = 0$ in the ansatz  (\ref{ansx}) and discarding the second (perturbative) part in the approximation
\beq
V_{\mathrm{C}}(p)\approx \frac{8 \pi \sigma_{\mathrm{C}}}{p^4} + \frac{g^2}{p^2} \label{foo}
\eeq
for the Coulomb potential [Eq.~(\ref{Gx})]. As one observes, the inclusion of the coupling to the transversal gluon changes only the mid- and large-momentum regime while the infrared behavior is not changed at all. This is perhaps a little bit surprising but should have been expected in view of the fact that the non-Abelian Coulomb term [the first part in eq.~(\ref{foo})], which gives rise to a linearly rising potential at large distances, dominates the infrared behavior of the gap equation. Let us also mention that we do not find chiral symmetry breaking from our equations when the linearly rising part of the Coulomb potential is neglected.

\section{Hamiltonian approach to finite temperature QCD by compactifying a spatial dimension}

In Refs.~\cite{Reinhardt2011, Heffner2012} the variational approach to Yang--Mills theory in Coulomb gauge was extended to finite temperatures by making a quasi-particle ansatz for the density matrix of the grand canonical ensemble where the quasi-particle energy was determined by minimizing the free energy. The resulting variational equations could be solved analogously to the ones at zero temperature. There is, however, a more efficient way to treat Yang--Mills theory at finite temperature within the Hamiltonian approach.
The motivation comes from the Polyakov loop
\beq
P[A_0](\vx) = \frac{1}{d_r} \tr \, \mathcal{P} \exp\left[\ii g \il_0^{L} \dd x^0 \, A_0 (x^0, \vx) \right] \, , \label{Gl: Polyakovloop}
\eeq
where $A_0 = A_0^a t^a$ is the temporal gauge field in the fundamental representation, $\mathcal{P}$ is the path ordering prescription, $d_r$ denotes the dimension of the representation of the gauge group and $L = 1/T$ is the length of the compactified Euclidean time axis which represents the inverse temperature. The Polyakov loop cannot be calculated straightforwardly in the Hamiltonian approach due to the unrestricted time interval and the use of the Weyl gauge $A_0 = 0$. Both problems are overcome in the more efficient Hamiltonian approach to finite-temperature quantum field theory developed in Ref.~\cite{Reinhardt:2016xci}. This approach does not require an ansatz for the density matrix of the grand canonical ensemble and allows the evaluation of the Polyakov loop. In this novel approach, one exploits the $O(4)$ invariance to interchange the Euclidean time axis with one spatial axis. The temporal (anti-)periodic boundary conditions to the fields become then spatial boundary conditions, while the new (Euclidean) time axis has infinite extent as is required within the Hamiltonian approach (see below). The upshot is that the partition function at finite temperature $L^{- 1}$ is entirely given by the ground state calculated on the spatial manifold $\R^2 \times S^1(L)$, where $S^1(L)$ is a circle with length $L$. The whole thermodynamics of the theory is then encoded in the vacuum calculated on the partially compactified spatial manifold $\R^2 \times S^1(L)$. This approach was used in Ref.~\cite{Heffner2015} to study Yang--Mills theory at finite temperature and in Ref.~\cite{RH2013} to calculate the Polyakov loop within the Hamiltonian approach. Furthermore, in Ref.~\cite{RV2016} the so-called dual quark condensate was evaluated using this approach. Let us briefly sketch its main properties: \\

Consider finite-temperature quantum field theory in the standard functional integral approach. Here the finite temperature is introduced by going to Euclidean space and compactifying the Euclidean time dimension by imposing periodic and antiperiodic boundary conditions for Bose and Fermi fields, respectively,
\begin{subequations} \label{656-32}
\begin{align}
A(x^0 = L/2) &= A(x^0 = -L/2) \, , \\
\psi(x^0 = L/2) &= -\psi(x^0 = - L/2) \, .
\end{align}
\end{subequations}
The length of the compactified dimension $L$ represents then the inverse temperature $T^{-1} = L$. One can now exploit the $O(4)$ invariance of the Euclidean Lagrangian to rotate the Euclidean time axis into a space axis and, correspondingly, one spatial axis into the Euclidean time axis. Of course, thereby all vectorial quantities transform in the same way, i.e.~we can choose the transformation:
\begin{alignat}{5}
\label{G43}
& x^0 &\to x^3 \, , &\quad \quad A^0 &\to A^3 \, , &\quad \quad \gamma^0 &\to \gamma^3 \, , \nonumber\\
& x^1 &\to x^0 \, , &\quad \quad A^1 &\to A^0 \, , &\quad \quad \gamma^1 &\to \gamma^0 \, .
\end{alignat}
After this rotation we are left with the spatial  periodic and antiperiodic boundary conditions
\begin{subequations} \label{G44}
\begin{align}
A(x^3 = L/2) &= A (x^3 = - L/2) \, , \nonumber\\
\psi(x^3 = L/2) &= - \psi (x^3 = - L/2) \, .
\end{align}
\end{subequations}
As a consequence of the $O(4)$ rotation our spatial manifold is now $\R^2 \times S^1(L)$ instead of $\R^3$ while the temporal manifold is $\R$ independent of the temperature, i.e.~the temperature is now encoded in one spatial dimension while time has infinite extension. We can now apply the usual canonical Hamiltonian approach to this rotated space-time manifold. As the new time axis has infinite extension 
$\ell \to \infty$, the partition function is now given by
\beq
Z(L) = \lim\limits_{\ell \to \infty} \mathrm{tr} \exp(- \ell H(L)) \, , \label{G45}
\eeq
where $H(L)$ is the usual Hamiltonian obtained after canonical quantization, however, now defined on the spatial manifold $\R^2 \times S^1(L)$. Taking the trace in the basis of the exact eigenstates of the Hamiltonian $H(L)$, we obtain for the partition function (\ref{G45})
\beq
Z(L) = \lim\limits_{\ell \to \infty} \sli_n \exp (- \ell E_n (L)) = \lim\limits_{\ell \to \infty} \exp (- \ell E_0 (L)) \, . \label{G46}
\eeq
The full partition function is now obtained from the ground state energy calculated on the spatial manifold $\R^2 \times S^1(L)$. Introducing the energy density $e(L)$ on $\R^2 \times S^1(L)$ by separating the volume $L \ell^2$  of the spatial manifold from the energy we have
\beq
E_0(L) = L \ell^2 e(L) \, . \label{G47}
\eeq
For the physical pressure
\beq
P = \frac{1}{L} \frac{\partial \ln Z}{\partial V} \, , \qquad V = \ell^3 \label{949-949}
\eeq
one finds from (\ref{G46})
\beq
P = - e(L) \, , \label{G48}
\eeq
while the physical energy density $\varepsilon$ is obtained as 
\beq
\varepsilon = \frac{\partial (L e (L))}{\partial L} - \mu \frac{\partial e (L)}{\partial \mu} \, . \label{G49}
\eeq
To distinguish this quantity from the (negative) Casimir pressure $e(L)$ [Eq.~(\ref{G48})], which also appears as an energy density in our formalism after the transformation [Eq.~(\ref{G43})], we will denote $e(L)$ as \textit{pseudo energy density}. Finally, after the $O(4)$ rotation, eq.~(\ref{G43}), the finite chemical potential $\mu$ enters the single-particle Dirac Hamiltonian $h$ in the form
\beq
h(\mu) = h(\mu=0) + \ii \mu \alpha^3 \, , \label{G50}
\eeq
where $\alpha^3$ is the third Dirac matrix and $h(\mu=0)$ the standard Dirac operator coupled to the gauge field.

\subsection{Free Bose and Fermi gases}

\begin{figure}
\centering
\begin{subfigure}{0.45\textwidth}
\includegraphics[width=\textwidth,clip]{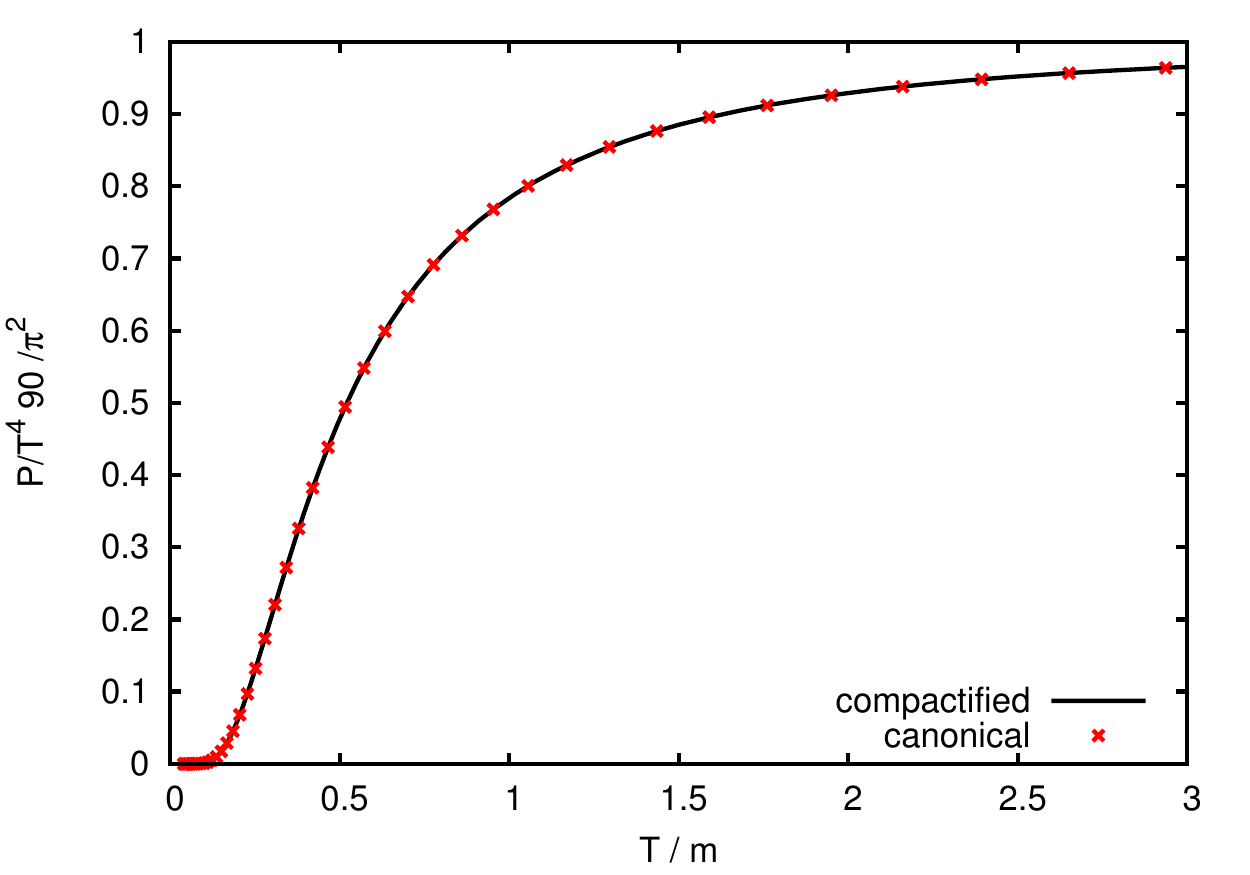}
\caption{}
\end{subfigure}
\quad
\begin{subfigure}{0.45\textwidth}
\includegraphics[width=\textwidth,clip]{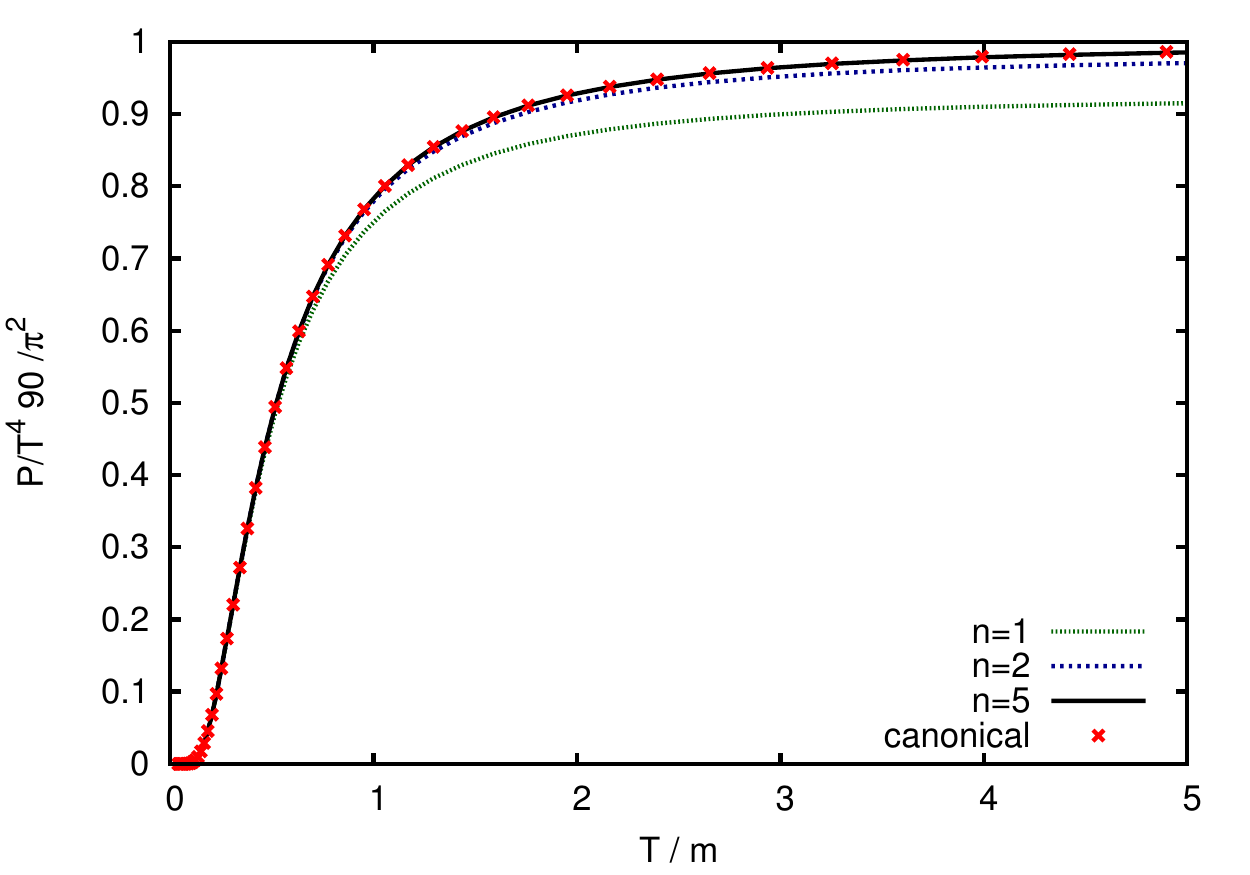}
\caption{}
\end{subfigure}
\caption{The pressure of a free massive Bose gas (a) calculated from eq.~(\ref{G55}) (full curve) and from the grand canonical ensemble (\ref{G51}) (crosses). (b) The pressure when the summation index in eq.~(\ref{G55}) is restricted to $|n| = 1, 2$ and $5$.}
\label{fig-10}
\end{figure}

To illustrate the above approach let us first consider a relativistic Bose gas with dispersion relation $\omega(p) = \sqrt{\vp^2 + m^2}$, where we assume for simplicity a vanishing chemical potential. The 
thermodynamical pressure obtained from the grand canonical ensemble for such a system is given by
\beq
P = \frac{2}{3} \int \frac{\dd^3 p}{(2 \pi)^3} \frac{p^2}{\omega (p)} n (p) \, , \quad \quad n (p) = \frac{1}{\exp(\beta \omega (p)) - 1} \, , \label{G51}
\eeq
where $n(p)$ are the finite temperature Bose occupation numbers. On the other hand,
one finds for the ideal Bose gas with dispersion relation $\omega (p) = \sqrt{\vp^2 + m^2}$ 
the pseudo energy density on the spatial manifold $\R^2 \times S^1(L)$ \cite{Reinhardt:2016xci}
\beq
e(L) = \frac{1}{2} \int \frac{\dd^2 p_\perp}{(2 \pi)^2} \frac{1}{L} \sli^\infty_{n = - \infty} \sqrt{\vp^2_\perp + {p_n}^2 + m^2} \, , \quad \qquad p_n = \frac{2 n \pi}{L} \, , \label{G52}
\eeq
where $p_n$ are the bosonic Matsubara frequencies. This quantity does not look at all like the negative of the pressure (\ref{G51}), as it should by eq.~(\ref{G48}). In fact, as it stands it is ill defined: the integral and the sum are both divergent. To make it mathematically well defined, we first use the proper-time regularization of the square root,
\beq
\sqrt{A} = \frac{1}{\Gamma \lk - \frac{1}{2} \rkx} \lim\limits_{\Lambda \to \infty}\left[\il^\infty_{1/\Lambda^2} \dd \tau \,\tau^{-\frac{1}{2}}\, \exp (- \tau A) - 2 \Lambda + \mathcal{O}(\Lambda^{-1})\right] \, . \label{G53}
\eeq
The divergent constant appears because the limit $\Lambda \to \infty$ of the incomplete $\Gamma$-function is not smooth; it drops out when taking the difference to the zero-temperature case after eq.~(\ref{G55}) below. With this replacement, the momentum integral in eq.~(\ref{G52}) can be carried out in closed form. For the remaining Matsubara sum we use the Poisson resummation formula,
\beq
\frac{1}{2 \pi} \sli^\infty_{k = -\infty} \exp(\ii k x) = \sli^\infty_{n = - \infty} \delta (x - 2 \pi n) , \label{G54}
\eeq
after which the proper-time integral can also be carried out, yielding for the pseudo energy density (\ref{G52}) 
\beq
e(L) = - \frac{1}{2 \pi^2} \sli^\infty_{n = -\infty} \lk \frac{m}{n L} \rkx^2 K_2(n L m) \, , \label{G55}
\eeq
where $K_\nu(z)$ is the modified Bessel function. The term with $n = 0$ is divergent and represents the pseudo energy density of the zero temperature vacuum, which should be eliminated from the pressure. The remaining terms $n \neq 0$ are all finite and also the remaining sum converges. This sum, however, cannot be carried out analytically for massive bosons (the same
applies to the integral in the grand canonical expression (\ref{G51}) for the pressure). In the zero-mass limit, we find from eq.~(\ref{G55}) for the pressure (\ref{G48})
\beq
P = \frac{\zeta(4)}{\pi^2} T^4 = \frac{\pi^2}{90} T^4 , \label{G56}
\eeq
which is Stefan--Boltzmann law, the correct result also obtained from the grand canonical ensemble. For massive bosons the evaluation of the sum in eq.~(\ref{G55}) as well as the evaluation of the integral in eq.~(\ref{G51}) have to be done numerically. The result is shown in fig.~\ref{fig-10} (a). As expected the pressure calculated from the compactified spatial dimension reproduces the result of  the usual grand canonical ensemble. Figure~\ref{fig-10} (b) shows the various contributions to the pressure. It is seen that only a few terms in the sum of eq.~(\ref{G55}) are necessary to reproduce the result of the grand canonical ensemble to good accuracy.

In the case of the relativistic \emph{Fermi} gas with dispersion relation $\omega (p) = \sqrt{\vp^2 + m^2}$ the energy density on $\R^2 \times S^1(L)$  is given by
\beq
e(L) = - 2 \int \frac{\dd^2 p_\perp}{(2 \pi)^2} \frac{1}{L} \sli^\infty_{n = - \infty} \sqrt{\vp^2_\perp + (p_n + \ii \mu)^2 + m^2} \, , \quad \quad p_n = \frac{2 n + 1}{L} \pi \, , \label{G57}
\eeq
where we have now included a non-vanishing chemical potential $\mu$. To make this expression mathematically well-defined one has to resort again to  the proper-time regularization and Poisson resummation technique sketched above. The result is
\beq
e(L) = \frac{2}{\pi^2} \sli^\infty_{n = 0} \cos \left[ n L \lk \frac{\pi}{L} - \ii \mu \rkx \right] \lk \frac{m}{n L} \rkx^2 K_{- 2} (n L m) \, . \label{G58}
\eeq
Again, the term with $n = 0$ represents the zero temperature vacuum energy density, which is divergent and has to be removed. As before, this expression can only be calculated in closed form for massless particles. For the remaining sum to converge, an analytic continuation $\ii \mu L  \to \bar{\mu} \in \R$ is required to carry out the sum
\beq
\sli^\infty_{n = 1} (-1)^n \frac{\cos (n \bar{\mu})}{n^4} = \frac{1}{48} \left[ - \frac{7}{15} \pi^2 + 2 \pi^2 \bar{\mu}^2 - \bar{\mu}^4 \right] \, . \label{G59}
\eeq
Continuing back to real chemical potentials one finds through eq.~(\ref{G48}) for the pressure
\beq
P = \frac{1}{12 \pi^2} \left[ \frac{7}{15} \pi^4 T^4 + 2 \pi^2 T^2 \mu^2 + \mu^4 \right] \, , \label{G60}
\eeq
which is the correct result obtained also from the usual grand canonical ensemble.

In Ref.~\cite{Heffner2015}, the above approach was used to study Yang--Mills theory at finite temperature. For this purpose, it is merely required to repeat the variational Hamiltonian approach on the spatial manifold $\R^2 \times S^1 (L)$.  Due to the one compactified spatial dimension the three-dimensional integral equations of the zero-temperature case are replaced by a set of two-dimensional integral equations distinguished by different Matsubara frequencies. Below, I will use this approach to calculate the effective potential of the Polyakov loop, the order parameter of confinement.

\subsection{The Polyakov loop}

Consider SU($N$) gauge theory at finite temperature, where the temperature is introduced by the usual periodic boundary condition in the temporal direction (\ref{656-32}). Gauge
transformations preserving this boundary conditions need to be periodic only up to an element $z$ of the center $Z(N)$ of the gauge group,
\beq
U(x^0 = L) = z U(x^0 = 0) \, , \quad \quad z \in Z (N) \, . \label{G61}
\eeq
Since there are $N$ center elements, this theory has a residual global $Z(N)$ symmetry, which remains after gauge fixing. However, there are quantities which are sensitive to such a $Z(N)$ symmetry transformation. The most prominent  example is the Polyakov loop (\ref{Gl: Polyakovloop}). A gauge transformation of the form (\ref{G61}) multiplies the Polyakov loop by the center element $z$, i.e.
\beq
P[A^U_0] = z P [A_0] \, . \label{G63}
\eeq
The expectation value of the Polyakov loop
\beq
\langle P[A_0](\vx) \rangle \sim \exp \lk - F_\infty(\vx) L \rkx \label{G64}
\eeq
can be shown to be related to the free energy $F_\infty(\vx)$ of a static color point charge located at $\vx$ \cite{Svetitsky1986}. In a confining theory this quantity has to be infinite since there are no free color charges, while in a deconfined phase it is finite. Accordingly we find for the expectation value of the Polyakov loop
\beq
\langle P [A_0] (\vx) \rangle \begin{cases}
                               = 0 &\quad \text{confined phase,} \\
                               \neq 0 &\quad \text{deconfined phase.}
                              \end{cases} \label{G65}
\eeq
From eq.~(\ref{G63}) follows that a state with vanishing expectation value of the Polyakov loop is invariant with respect to the global center transformation, while in the deconfined phase the $Z(N)$ center symmetry is obviously broken. In the continuum theory the Polyakov loop can be most easily calculated in the Polyakov gauge
\beq
\partial_0 A_0 = 0 , \qquad A_0 \text{ color diagonal.} \label{G66}
\eeq
In this gauge one finds, for example, for the SU(2) gauge group that the Polyakov loop
\beq
P [A_0] (\vx) = \cos \lk \frac{1}{2} g A_0 (\vx) L \rkx \, \label{G67}
\eeq
is a one-to-one function of the gauge field, at least in the fundamental modular region of this gauge. It can be shown, see Refs.~\cite{BGP2010, MP2008}, that instead of the expectation value of the Polyakov loop $\langle P[A_0] \rangle$ one may alternatively use the Polyakov loop of the expectation value, $P [\langle A_0 \rangle ]$, or the expectation value of the temporal gauge field itself, $\langle A_0 \rangle$, as order parameter of confinement in the gauge (\ref{G66}). This analysis also shows that the most efficient way to obtain the Polyakov loop is to carry out a so-called background field calculation with a temporal background field $a_0 (\vx) = \langle A_0 (\vx) \rangle$ chosen in the Polyakov gauge, and then calculate the effective potential $e [a_0]$ of that background field. From the minimum $\bar{a}_0$ of this potential one evaluates the Polyakov loop $P[\langle A_0\rangle] = P[\bar{a}_0]$, which can then serve as the order parameter of confinement.

\begin{figure}
\centering
\begin{subfigure}{0.45\textwidth}
\includegraphics[width=\textwidth,clip]{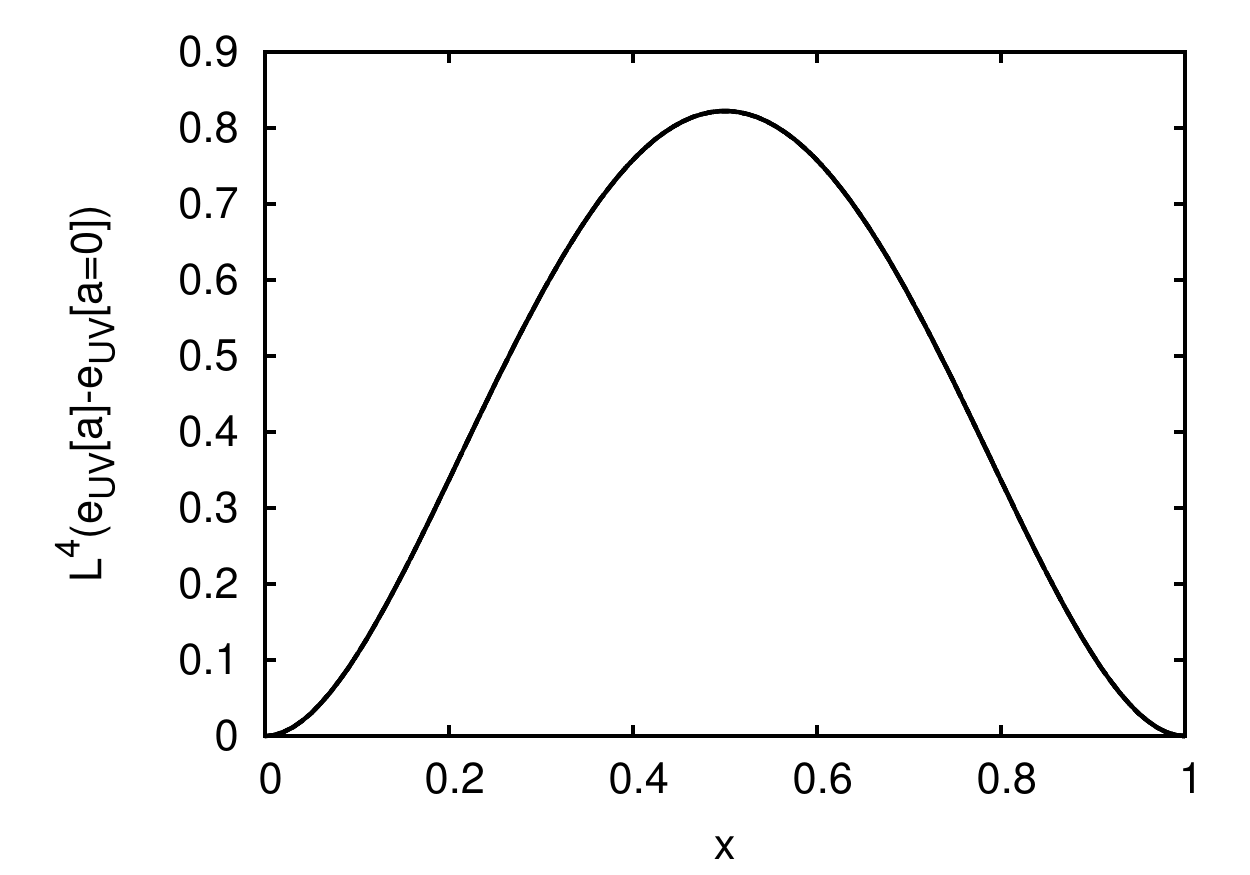}
\caption{}
\end{subfigure}
\quad
\begin{subfigure}{0.45\textwidth}
\includegraphics[width=\textwidth,clip]{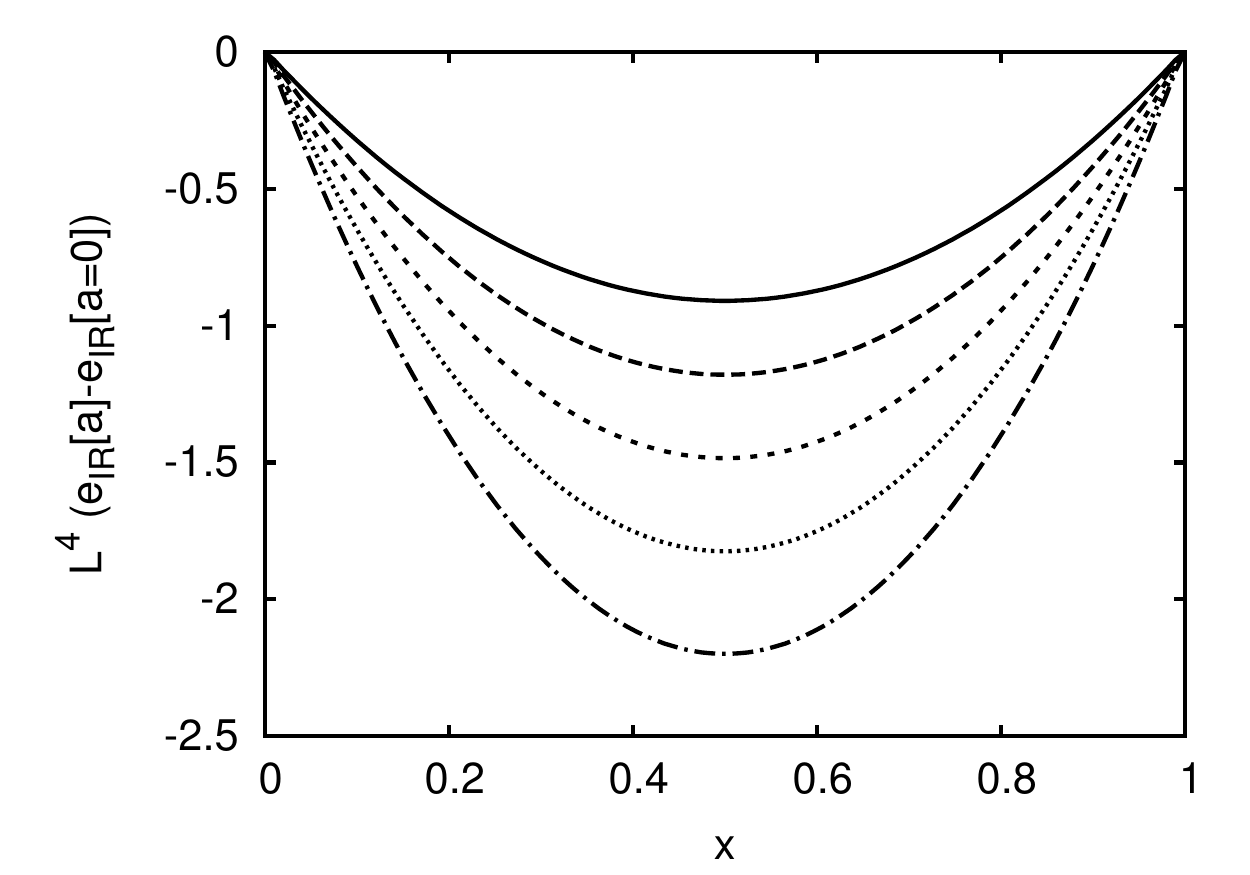}
\caption{}
\end{subfigure}
\caption{The effective potential of the Polyakov loop $e(a, L)$ (\ref{G68}) as function of the background field $x = a_3 L / 2 \pi$. The curvature is neglected $(\chi = 0)$ and the gluon energy assumed to be (a) $\omega (p) = p$ (UV-form) and (b) $\omega (p) = M^2 / p$ (IR-form), respectively.}
\label{fig-11}
\end{figure}

Such a calculation was done a long time ago in Ref.~\cite{Weiss:1980rj,*Gross:1980br}, where the effective potential $e[a_0]$ was calculated in one-loop perturbation theory. The result is shown in fig.~\ref{fig-11} (a). The potential is periodic due to center symmetry. The minimum of the potential occurs at the vanishing background field, which gives $P[a_0 = 0] = 1$ corresponding to the deconfined phase. This is, of course, expected due to the use of perturbation theory. Below, I present the results of a non-perturbative evaluation of $e[a_0]$ in the Hamiltonian approach in Coulomb gauge.

At first sight it seems that the Polyakov loop cannot be calculated in the Hamiltonian approach due to the use of the Weyl gauge $A_0 = 0$. However, we can now use the alternative Hamiltonian approach to finite temperature introduced above, where the temperature is introduced by compactifying a spatial dimension. Here, we compactify the $x_3$-axis and consequently put also the background field along this axis, $\va = a \ve_3$. In the Hamiltonian approach the effective potential of a spatial background field $\va$ can be easily calculated by minimizing the expectation value of the Hamiltonian under the constraint $\langle \vA \rangle = \va$. The resulting energy $\langle H \rangle_{\va} = L^2 \ell e(\va)$ is then (up to the spatial volume factor) the effective potential. So the effective potential $e(\va)$ is nothing but the pseudo energy density considered earlier, but now calculated in a background gauge with the contraint $\langle \vA \rangle = \va$.

\subsection{The effective potential of the Polyakov loop}

After lengthy calculations exploiting the gluon gap equation (\ref{G8}), one finds for the effective potential of the Polyakov loop the following expression
\beq
e(a, L) = \sli_\sigma \frac{1}{L} \sli^\infty_{n = - \infty} \int \frac{\dd^2 p_\perp}{(2 \pi)^2} \lk \omega (\vp^\sigma) - \chi (\vp^\sigma) \rkx \, , \label{G68}
\eeq
where $\omega(p)$ is the gluon energy and $\chi(p)$ is the ghost loop. These quantities have to be taken with the momentum variable
\beq
\vp^\sigma = \vp_\perp + \lk p_n - \sigma \cdot a \rkx \ve_3 \, , \label{G69}
\eeq
where $\vp_\perp$  is the momentum corresponding to the two uncompactified space dimensions while $p_n = 2 \pi n / L$ is the Matsubara frequency resulting from the compactification of the third dimension. Furthermore, $\sigma \cdot a \equiv \sigma^b a^b$ denotes the product of the color background field with the root vectors $\sigma^b$ of the gauge group. Equation~(\ref{G68}) includes also the summation over the roots $\sigma$ of the gauge group. In Refs.~\cite{Reinhardt2012, RH2013}, the effective potential (\ref{G68}) was explicitly calculated using for $\omega (p)$ and $\chi (p)$ the results from the variational calculation in Coulomb gauge at zero temperature \cite{ERS2007}. This represents certainly an approximation since, in principle, one should use the finite-temperature solutions obtained in Ref.~\cite{Heffner2015}.

\begin{figure}
\centering
\begin{subfigure}{0.45\textwidth}
\includegraphics[width=\textwidth,clip]{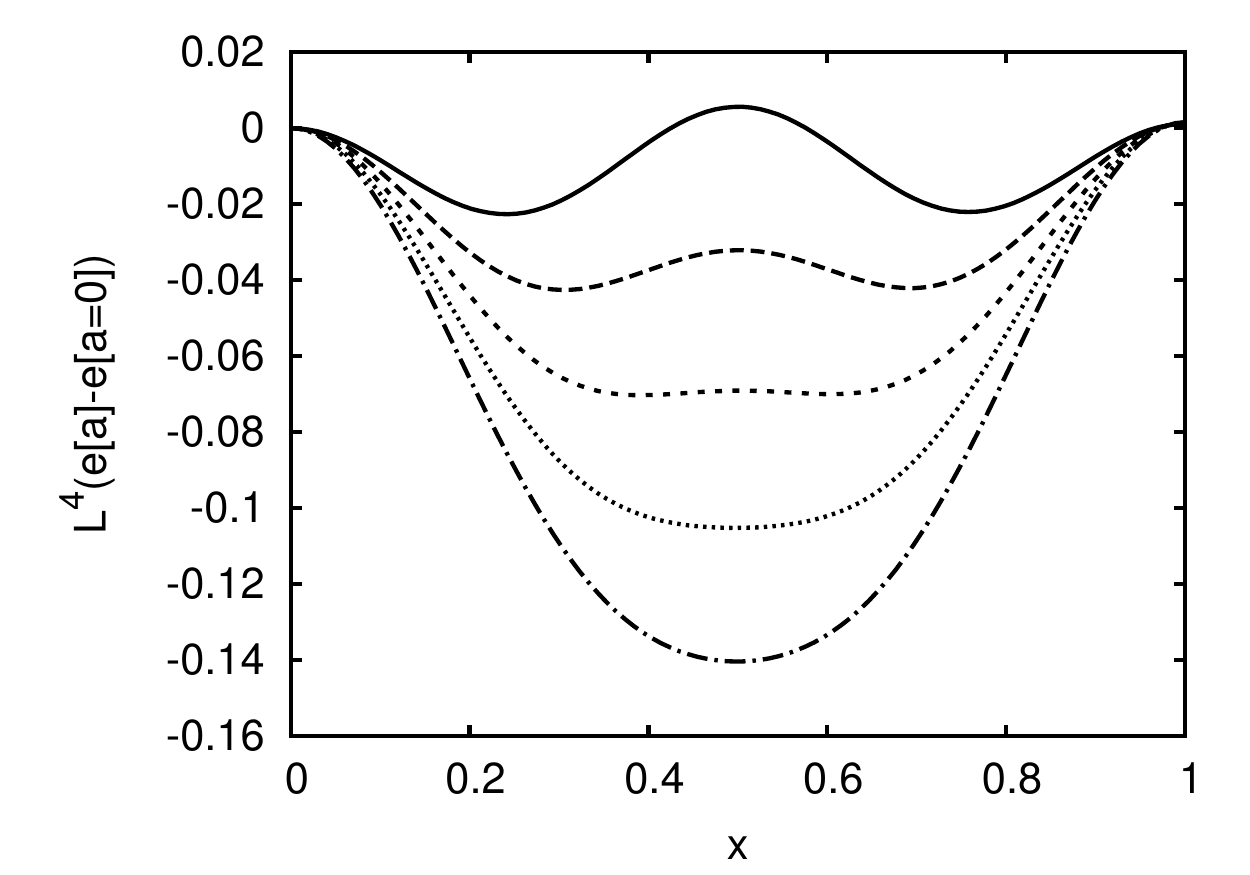}
\caption{}
\end{subfigure}
\quad
\begin{subfigure}{0.45\textwidth}
\includegraphics[width=\textwidth,clip]{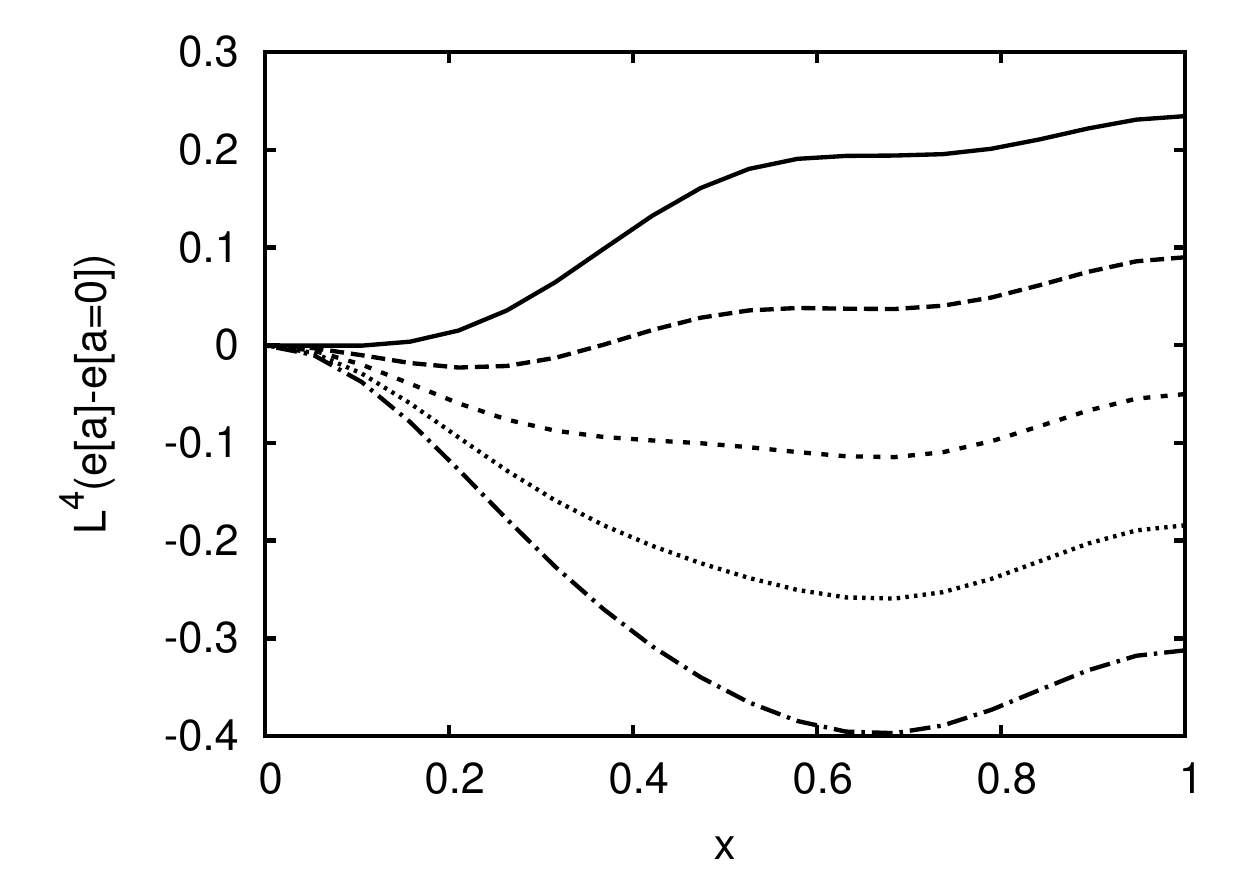}
\caption{}
\end{subfigure}
\caption{Effective potential of the Polyakov loop (\ref{G68}) as function of the background field $x = a_3 L / 2 \pi$ at various temperatures, for the gauge group (a) SU(2) and (b) SU(3).}
\label{fig-12}
\end{figure}

\begin{figure}
\centering
\begin{subfigure}{0.45\textwidth}
\includegraphics[width=\textwidth,clip]{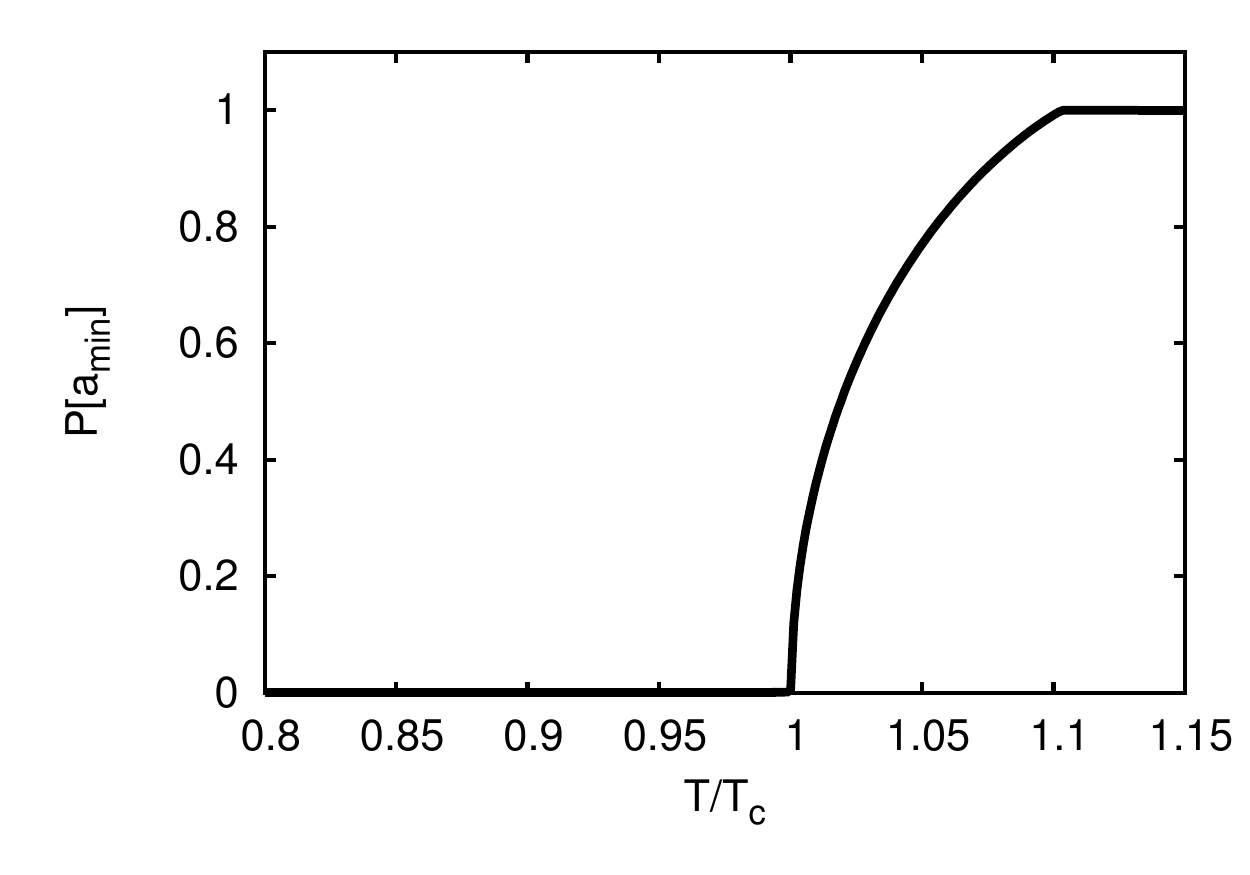}
\caption{}
\end{subfigure}
\quad
\begin{subfigure}{0.45\textwidth}
\includegraphics[width=\textwidth,clip]{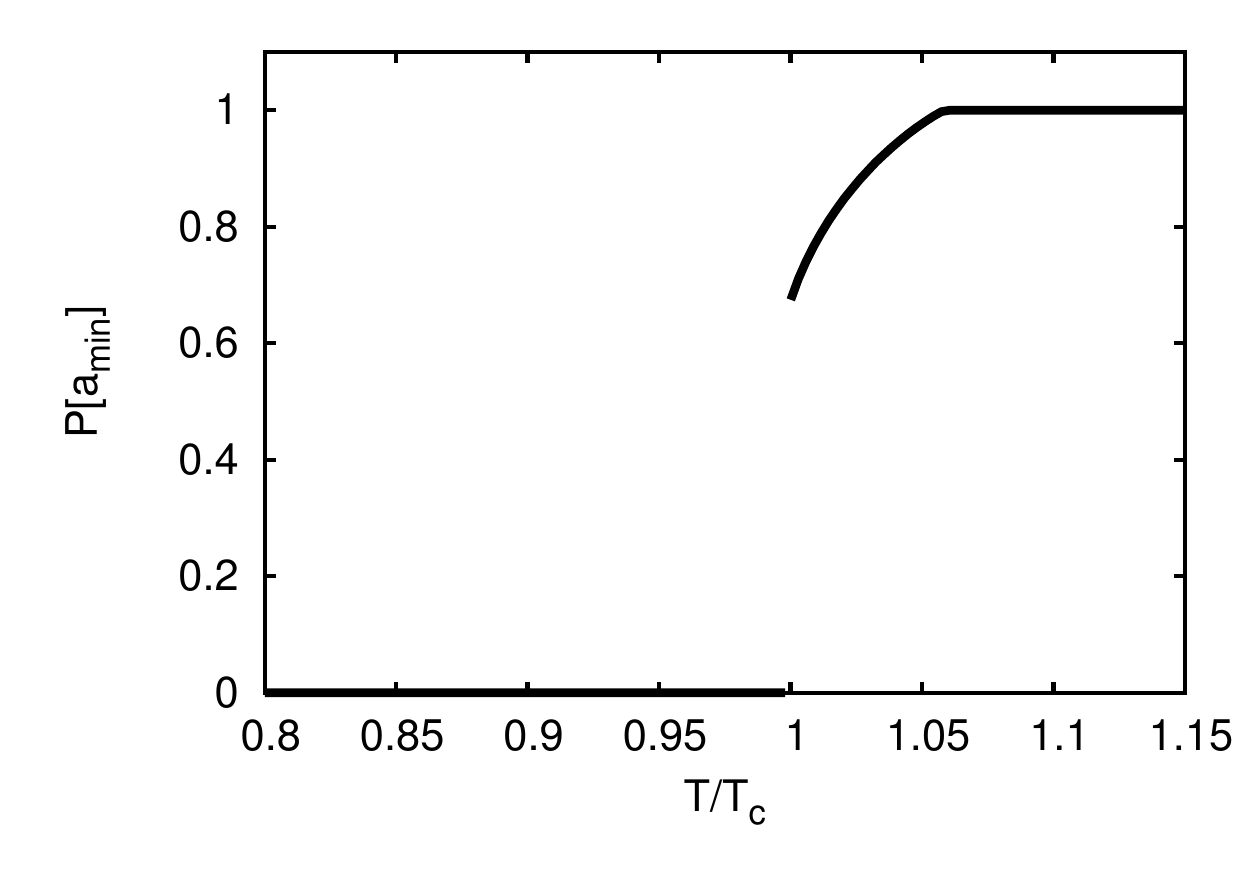}
\caption{}
\end{subfigure}
\caption{The Polyakov loop as function of the temperature (a) for SU(2) and (b) for SU(3).}
\label{fig-13}
\end{figure}

Before I present the full results let me ignore the ghost loop $\chi(p)$ in eq.~(\ref{G68}) and consider the ultraviolet and infrared limit of the gluon energy. If we choose the ultraviolet limit $\omega(p) = p$, we obtain from eq.~(\ref{G68}) with $\chi(p) = 0$ precisely the Weiss potential, shown in fig.~\ref{fig-11} (a), which corresponds to the deconfined phase. Choosing for the gluon energy its infrared limit $\omega(p) = M^2 / p$, one finds from eq.~(\ref{G68}) with $\chi(p) = 0$ the (center symmetric) potential shown in fig.~\ref{fig-11} (b). From its center symmetric minimum $\bar{a} = \pi / L$ one finds a vanishing Polyakov loop $P[\bar{a}] = 0$ corresponding to the confined phase. Obviously, the deconfining phase transition results from the interplay between the confining infrared and the deconfining ultraviolet potentials. Choosing for the gluon energy the sum of the UV- and IR-parts $\omega(p) = p + M^2/p$, which can be considered as an approximation to the Gribov formula (\ref{G13}), one has to add the UV and IR potentials and finds a phase transition at a critical temperature $T_{\mathrm{c}} = \sqrt{3} M / \pi$. With the Gribov mass $M \approx 880 \, \mathrm{MeV}$ this gives a critical value of $T_{\mathrm{c}} \approx 485 \, \mathrm{MeV}$ for the color group SU(2), which is much too high as compared to the lattice value of $312\,\mathrm{MeV}$ \cite{Lucini:2003zr}. One can show analytically \cite{Reinhardt2012,RH2013} that the neglect of the ghost loop $\chi(p) = 0$ shifts the critical temperature to higher values. If one uses for the gluon energy $\omega(p)$ the Gribov formula (\ref{G13}) and includes the ghost loop $\chi(p)$, one finds the effective potential shown in fig.~\ref{fig-12} (a), which shows a second order phase transition and gives a transition temperature of $T_{\mathrm{c}} \approx 269 \, \mathrm{MeV}$ for the gauge group SU(2), which is in the right ballpark. The Polyakov loop $P[\bar{a}]$ calculated from the minimum $\bar{a}$ of the effective potential $e(a, L)$ (\ref{G68}) is plotted in fig.~\ref{fig-13}(a) as function of the temperature.
\begin{figure}
\centering
\begin{subfigure}{0.45\textwidth}
\includegraphics[width=\textwidth,clip]{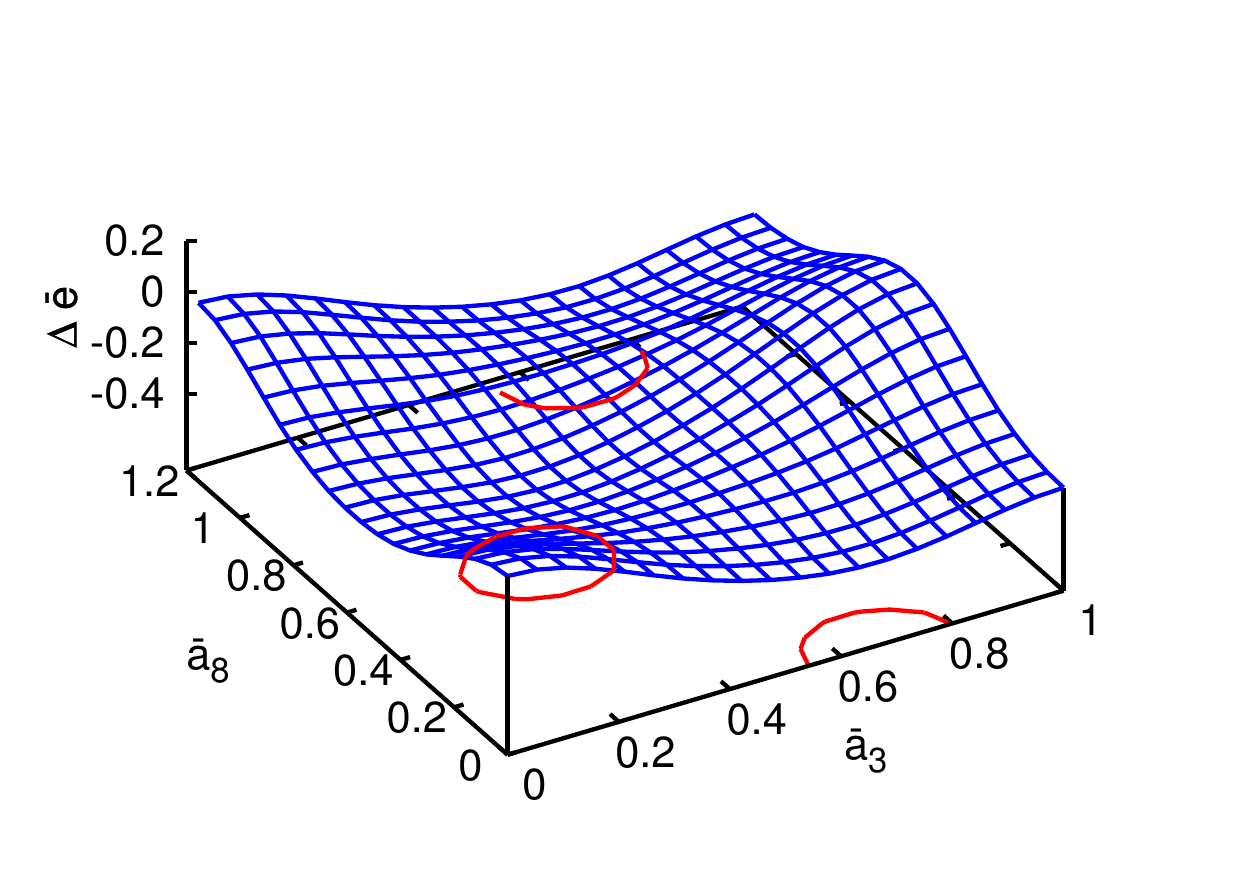}
\caption{}
\end{subfigure}
\quad
\begin{subfigure}{0.45\textwidth}
\includegraphics[width=\textwidth,clip]{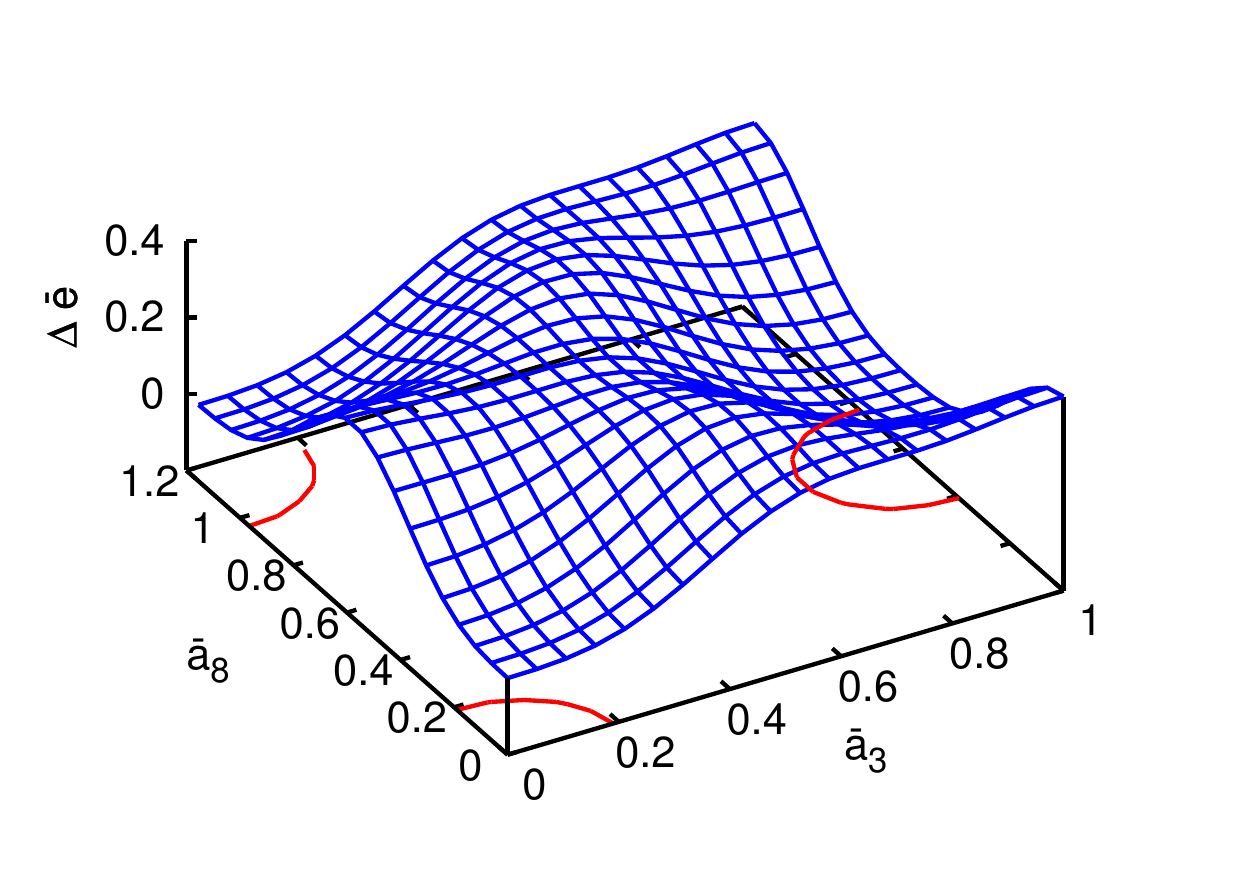}
\caption{}
\end{subfigure}
\caption{The effective potential of the Polyakov loop for the gauge group SU(3) as function of the two Cartan components of the background field $x = a_3 L / 2 \pi$ and $y = a_8 L / 2 \pi$ for (a) $T < T_{\mathrm{c}}$ and (b) $T > T_{\mathrm{c}}$.}
\label{fig-14}
\end{figure}

The effective potential for the gauge group SU(3) can be reduced to that of the SU(2) group by noticing that the SU(3) algebra consists of three SU(2) subalgebras characterized by the three positive roots $\vec{\sigma} = (1, 0)$, $(1/2, \sqrt{3}/2)$, $(1/2 \, , - \sqrt{3} /2)$. One finds 
\beq
e_{\mathrm{SU(3)}} (a, L) = \sum_{\boldsymbol{\sigma} > 0} e_{\mathrm{SU(2)}}[\vec{\sigma}] (a, L) \, . \label{931-70}
\eeq
The resulting effective potential for SU(3) is shown in fig.~\ref{fig-14} as function of the components of the background field in the Cartan algebra, $a_3$ and $a_8$. Above and below $T_{\mathrm{c}}$ the absolute minima of the potential occur in both cases for $a_8 = 0$. Cutting the two-dimensional potential surface at $a_8 = 0$, one finds the effective potential shown in fig.~\ref{fig-12} (b), which shows a first order phase transition with a critical temperature of $T_{\mathrm{c}} \approx 283 \, \mathrm{MeV}$. The first order nature of the SU(3) phase transition is also seen in fig.~\ref{fig-13} (b), where the Polyakov loop $P[\bar{a}]$ is shown as function of the temperature.

\subsection{The dual quark condensate}

\begin{figure}[ht]
\centering
\begin{subfigure}{0.45\textwidth}
\includegraphics[width=\textwidth,clip]{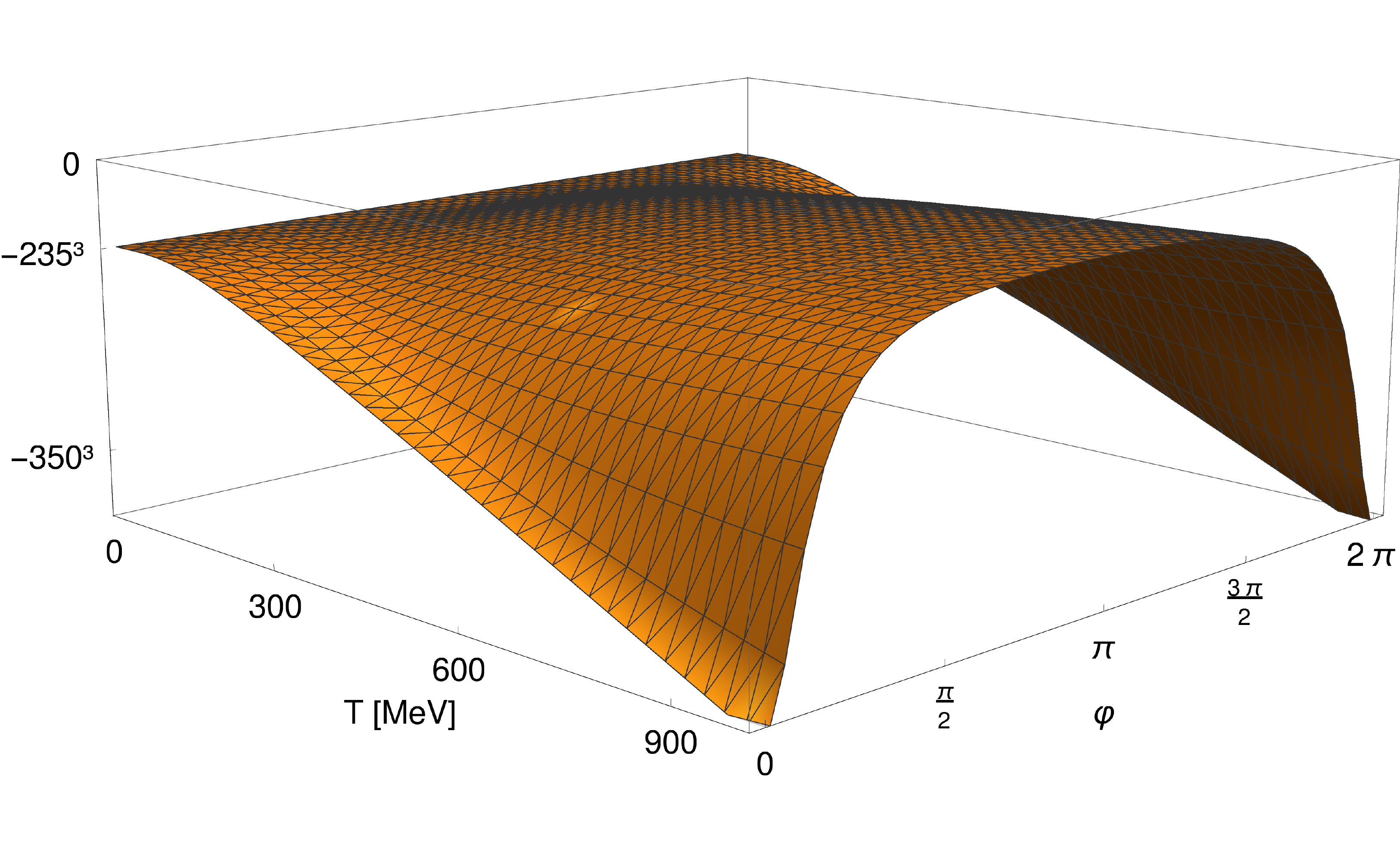}
\caption{}
\end{subfigure}
\quad
\begin{subfigure}{0.45\textwidth}
\includegraphics[width=\textwidth,clip]{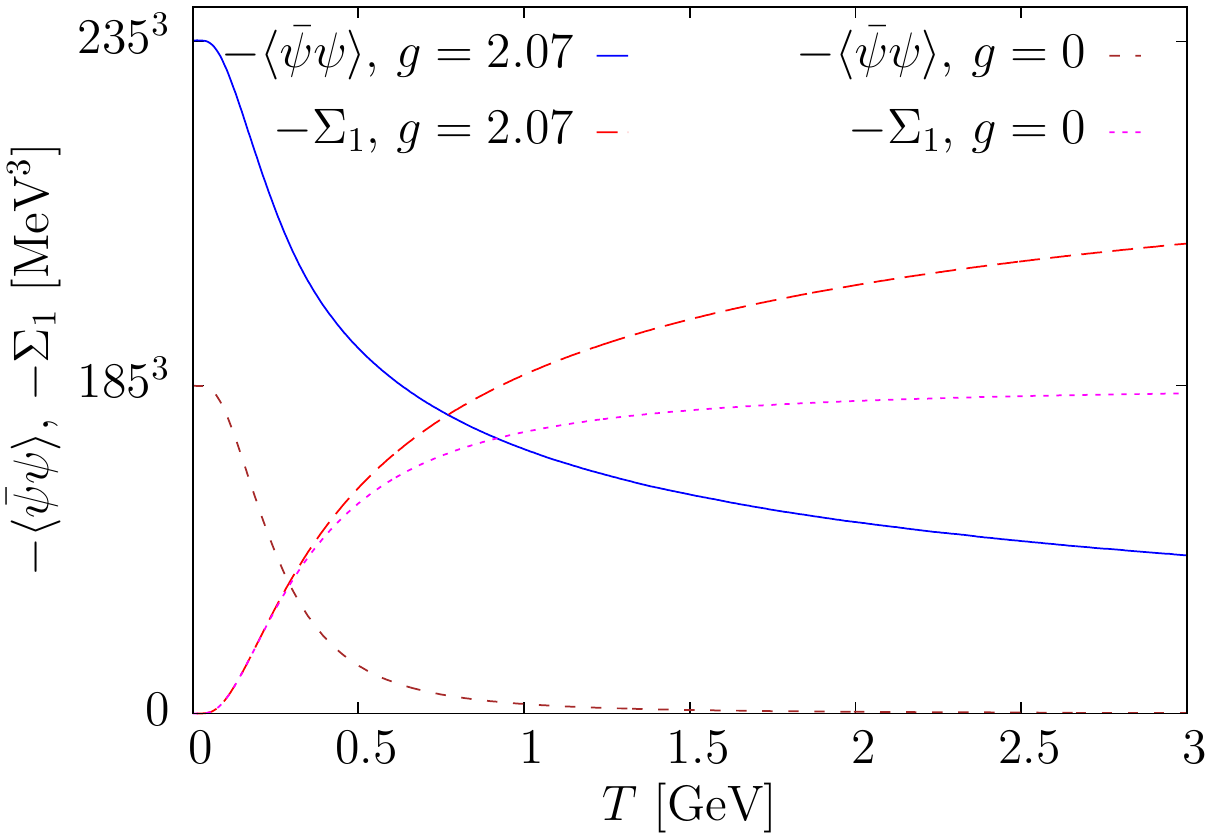}
\caption{}
\end{subfigure}
\caption{(a) Chiral quark condensate $\langle \bar{\psi} \psi \rangle_\varphi$ as function of the temperature $T$ and the phase $\varphi$ of the boundary condition (\ref{28}). (b) Chiral and dual quark condensate as function of the temperature. Results are presented for both a coupling of $g \simeq 2.1$ and $g = 0$.}
\label{fig6}%
\end{figure}

The dual quark condensate was originally introduced in Ref.~\cite{Gattringer2006} and was discussed in a more general context in Ref.~\cite{Synatschke:2007bz}. This quantity has been calculated on the lattice \cite{Bilgici:2008qy, Zhang2011}, in the functional renormalization group approach \cite{Braun:2009gm} and in the Dyson--Schwinger approach \cite{FMM2010}. The dual condensate is defined by
\beq
\Sigma_n = \int\limits_0^{2 \pi} \frac{\dd \varphi}{2 \pi} \exp(-\ii n \varphi) \langle \bar{\psi} \psi \rangle_\varphi \, , \label{27}
\eeq
where $\langle \bar{\psi} \psi \rangle_\varphi$ is the quark condensate calculated with the $U(1)$-valued boundary condition
\beq
\psi(x^0 + L/2, \vx) = \mathrm{e}^{\ii \varphi} \psi(x^0 - L/2, \vx) \, . \label{28}
\eeq
For $\varphi = \pi$ these boundary conditions reduce to the usual finite-temperature boundary conditions of the quark field in the functional integral representation of the partition function, see eq.~(\ref{656-32}). On the lattice it is not difficult to show that the quantity $\Sigma_n$ (\ref{27}) represents the vacuum expectation value of the sum of all closed Wilson loops winding precisely $n$-times around the compactified time axis. In particular, the quantity $\Sigma_1$ represents the expectation value of all closed loops winding precisely once around the compactified time axis and is therefore called the dressed Polyakov loop. The phase in the boundary condition (\ref{28}) can be absorbed into an imaginary chemical potential
\beq
\mu = \ii \frac{\pi - \varphi}{L} \label{29}
\eeq
for fermion fields satisfying the usual antisymmetric boundary condition $\psi(x^0 + L/2, \vx) = -\psi(x^0 - L/2, \vx)$. In the Hamiltonian approach to finite temperatures of Ref.~\cite{Reinhardt:2016xci}, where the compactified time axis has become the third spatial axis, the phase dependent boundary condition (\ref{28}) or equivalently the imaginary chemical potential (\ref{29}) manifests itself in the momentum variable along the (compactified) three-axis, which reads
\beq
p_3 = p_n + \ii \mu = \frac{2 \pi n + \varphi}{L} \, , \quad \quad p_n = \frac{2 n + 1}{L} \pi \, , \label{30}
\eeq
where $p_n$ is the usual fermionic Matsubara frequency [see eq.~(\ref{G57})]. Using the zero-temperature quark mass function $M(p)$ calculated in Ref.~\cite{QCDT0Rev}, one finds in the Hamiltonian approach to QCD of Ref.~\cite{QCDT0} for the dual quark condensate after Poisson resummation the leading expression \cite{RV2016}
\beq
\Sigma_n = -\frac{N}{\pi^2} \int\limits_0^{\infty} \dd p \, \frac{p^2 M(p)}{\sqrt{p^2 + M^2(p)}} \left[\delta_{n 0} + \frac{\sin(n \beta p)}{n \beta p}\right] \, , \label{31}
\eeq
where $N$ denotes the number of colors. In the same way, one can compute the quark condensate $\langle \bar{\psi} \psi \rangle_\varphi$ shown in fig.~\ref{fig6} (a). For the dressed Polyakov loop one finds the temperature behavior shown in fig.~\ref{fig6} (b), where we also compare with the result obtained when the coupling to the transversal gauge field degrees of freedom is neglected ($g = 0$). As one observes there is no difference at small temperatures in accord with the fact that the mass function $M(p)$ has the same infrared behavior, whether the coupling to the transversal gluons is included or not. The slower UV decrease of the full mass function causes the dual condensate to reach its high-temperature limit
\beq
\lim_{L \to 0} \Sigma_1 = -\frac{N}{\pi^2} \int\limits_0^{\infty} \dd p \, \frac{p^2 M(p)}{\sqrt{p^2 + M^2(p)}} = \lim_{L \to \infty} \langle \bar{\psi} \psi \rangle_{\varphi = \pi} \label{32}
\eeq
only very slowly. We expect, however, that this limit is reached faster when the finite-temperature solutions are used. This will presumably also convert the crossover obtained for the chiral condensate, see fig.~\ref{fig6} (b), into a true phase transition as expected for chiral quarks. From the inflexion points of the chiral and dual condensate one extracts the values of $T_{\chi}^{\mathrm{pc}} \simeq 170 \, \mathrm{MeV}$ and $T_{\mathrm{c}}^{\mathrm{pc}} \simeq 198 \, \mathrm{MeV}$ for the pseudo-critical temperatures of the chiral and deconfinement transition, respectively. For comparison, one finds
on the lattice for realistic quark masses $T_{\chi}^{\mathrm{pc}} \simeq 155 \, \mathrm{MeV}$ and $T_{\mathrm{c}}^{\mathrm{pc}} \simeq 165 \, \mathrm{MeV}$ \cite{Borsanyi2010, Bazavov2012}.

\section{Conclusions}

In my talk I have presented some recent results obtained within the Hamiltonian approach to QCD in Coulomb gauge. I have first shown that the so-called Coulomb string tension is not 
related to the temporal but to the spatial string tension. This relation explains the finite-temperature behavior of the Coulomb string tension, namely the fact that it does not disappear but even increases above the deconfinement transition. I have then studied the quark sector of QCD in Coulomb gauge using a Slater determinant ansatz for the quark wave functional, which includes in particular the quark-gluon coupling by two different Dirac structures. Our calculations show that there is no spontaneous breaking of chiral symmetry when the (linearly rising) infrared part of the Coulomb potential is excluded. Furthermore, choosing the Coulomb string tension from the lattice data we can reproduce the phenomenological value of the quark condensate when the coupling of the quarks to the transverse gluon is included.

I have then extended the Hamiltonian approach to QCD in Coulomb gauge to finite temperatures by compactifying a spatial dimension. Within this approach, I have calculated the effective potential of the Polyakov loop as well as the chiral and dual quark condensate as function of the temperature. Using our zero-temperature solution as input, from the Polyakov loop we predict a critical temperature for the deconfinement phase transition of about $T_{\mathrm{c}} \sim 275 \, \mathrm{MeV}$ for SU(2), and $T_{\mathrm{c}} \sim 280 \, \mathrm{MeV}$ for SU(3). Furthermore, the correct order of the phase transition was found for SU(2) and SU(3). For full QCD our calculations of the dual and chiral quark condensate
predict pseudo-critical temperatures of $T_{\chi}^{\mathrm{pc}} \simeq 170 \, \mathrm{MeV}$ for the chiral and $T_{\mathrm{c}}^{\mathrm{pc}} \simeq 198 \, \mathrm{MeV}$ for the deconfinement transition. In all these finite-temperature calculations the zero-temperature variational solutions were used as input, which is likely the reason that the critical temperatures currently obtained are too high as compared to lattice data. The solution of the variational principle at finite temperature will be the next step in our investigation of the QCD phase diagram.

\section*{Acknowledgement}

This work was supported in part by DFG-RE856/9-2 and by DFG-RE856/10-1.

\bibliography{QCDT0_nizza}

\end{document}